\documentclass[twocolumn]{aastex63}

\usepackage[utf8]{inputenc}
\usepackage{xcolor}

\usepackage{comment}
 \usepackage{hyperref}
\usepackage{amsmath}
 \usepackage{xfrac}

\newcommand\GALEX{{\it GALEX}}

 \accepted{April 20, 2022}

\submitjournal{PASP}  

\shorttitle{The FUV background}
 \shortauthors{Kulkarni}
\begin{document}

\title{The far ultra-violet background}

\correspondingauthor{S.\ R.\ Kulkarni}
 \email{srk@astro.caltech.edu}

\author[0000-0001-5390-8563]{S.\ R.\ Kulkarni}
 \affiliation{Owens Valley Radio Observatory 249-17, Caltech\\
Pasadena, CA 91125, USA}

\begin{abstract} The diffuse far-ultraviolet (FUV) background has
received considerable attention from astronomers since the
1970's.
The initial impetus came from the hope of detecting UV radiation
from the hot intergalactic medium. The central importance of the
FUV background to the physics (heating and ionization) of the diffuse
atomic phases motivated the next generation of experiments.  The
consensus view is that the diffuse FUV emission at high latitudes
has three components: stellar FUV reflected by dust grains (diffuse
Galactic light or DGL), FUV from other galaxies
 and the intergalactic medium
(extra-galactic background light or EBL) and a component of unknown
origin (and referred to as the ``offset" component).  During the1980's, 
there was some discussion that decaying dark matter particles
produced FUV radiation.  In this paper I  investigate production
of FUV photons by conventional sources: line emission from Galactic
Hot Ionized Medium,  two-photon emission from the Galactic Warm
Ionized Medium and low-velocity shocks, and Lyman-$\beta$ fluorescence
of hydrogen at several locales in the Solar System (the interplanetary
medium, the exosphere and the thermosphere of Earth). I conclude
that two thirds  and arguably all of the offset component can be
explained by the sum of the radiation from the processes listed
above.  
\end{abstract}

%\keywords{instrumentation: photometric and spectrographs --- surveys
 %--- supernovae: general --- catalogs}

\section{Background}
 \label{sec:Background}

The diffuse background in the wavelength range 912--2000\,\AA\ is
of central importance to the physics of the diffuse atomic phases
-- the Cold Neutral Medium (CNM) and the Warm Neutral Medium (WNM).
Even though historically, the various bands of ultra-violet (UV)
were defined by wavelength properties of detectors and mirror
coatings, here, following the Galaxy Evolution Explorer mission
(\GALEX), the Far-Ultraviolet (FUV) band
covers the wavelength range 1350--1750\,\AA.

The diffuse FUV radiation,  via photo-electric ionization of C~I
and photo-electric heating of dust particles, heats the two atomic
phases mentioned above.  The  same radiation, via ionization of
heavier elements  also provides residual ionization to the two
atomic phases.    FUV photons excite the Lyman and Werner bands of
H$_2$ \citep{dw80} which result in UV fluorescent line emission
and once in ten times leads to dissociations \citep{J74,db96}. In
the process, the strength of FUV radiation determines the transition
in diffuse clouds between atomic and molecular phases.

In the late sixties it was speculated that the hot intergalactic
medium (IGM) would be revealed by diffuse FUV emission (see
\citealt{ks70,pj80,H91} for summaries of missions).  Separately,
during the
 {1970's},
the first pulsar surveys found that pulsars signals were invariably
dispersed (see, for example, \citealt{mt77}).  During the {1980's}, thanks primarily to the work of Ronald J.\ Reynolds,
the same ionized medium was complementarily sensed via H$\alpha$
recombination emission. Together these two approaches established
the Warm Ionized Medium (WIM) as a distinct phase of the ISM (see
\citealt{hdb+09} for a review).

The filling factor of the WIM, by volume, is estimated to be between
20\% and 40\% of the Galactic disk. The required large ionizing
flux is conventionally attributed to Lyman continuum leaking from
OB associations \citep{R90b}.  The observed large vertical scale
height, $\approx 1\,$kpc, of the WIM was initially a mystery. This
``crisis" led to a resurrection of a hypothesis of decaying dark
matter as a major source of ionizing photons (\citealt{S80};
\citealt{S90}; more recent references include \citealt{kwo+14,hmo+15}).
\citet{mc93} and \citet{ds94} provided a conventional explanation
based on O stars as the principal source of ionizing radiation,
radiative transfer modeling and clustering of star-forming regions,
``chimneys" and ``channels" while \citet{smh00} made a strong case
for (high latitude) ionizing radiation resulting from old supernova
remnants.

Two missions, both launched in 2003, greatly advanced the field of
diffuse FUV radiation. \GALEX\ \citep{mfs+05} carrying both an FUV
and a near-UV  (NUV; 1750--2800\,\AA) wide-field imager, each with
a field-of-view (FoV)  of over a square degree but with pixel size
of only a few arc-seconds, undertook an FUV \&\ NUV sky survey.
STSAT-1 carrying the Far-UV Imaging Spectrometer (FIMS\footnote{also
sometimes referred to as SPEAR}; \citealt{eka+06}) aimed to study
the Galactic Hot Ionized Medium (HIM) by undertaking spectral
(900--1750\,\AA) imaging of the sky at few arc-minute spatial
resolution.

The overall consensus is that much of the diffuse FUV emission is
due to  reflection of stellar FUV photons by diffuse (``cirrus")
clouds and conveniently traced by IRAS $100\,\mu$m band or fluorescence
by molecular hydrogen.  Together this emission is called as the
Diffuse Galactic Light (DGL).  However, some diffuse background
emission is {\it not} correlated with cirrus clouds, a fraction of
which can be reasonably attributed to collective emission from other
galaxies and the intergalactic medium -- the so-called Extragalactic Background Light (EBL).
There remains some 120--180\,CU emission in the FUV band which
cannot be attributed to either DGL or EBL and was given the name
``offset" component.\footnote{DGL is proportional to the column
density of dust. However,  this linear relation has an offset:
radiation even when the dust column is very low -- hence the name
``offset".} Here,  CU (``continuum unit") stands for ${\rm
photon\,cm^{-2}\,s^{-1}}\,\text{\AA}^{-1}\,{\rm sr^{-1}}$  with
``sr" as a short for steradian.  A related unit is ``line unit"
(LU) which stands for ${\rm photon\,cm^{-2}\,s^{-1}\,sr^{-1}}$.  We
take the occasion to introduce ``Rayleigh" ($R$) which is a unit
for surface brightness of line emission and routinely used in
aeronomy.  Numerically, one Rayleigh is $10^6/(4\pi)\,{\rm
photon\,cm^{-2}\,s^{-1}\,sr^{-1}}$.

Separately, the irreducible background for an UV--IR
mission located at or about 1\,AU, is due to scattering of solar
photons by zodiacal dust. However, the Sun is faint in the FUV.
Ergo, the sky is dark in the FUV band. As a result, the FUV band
is most attractive for low surface brightness imaging of galaxies
\citep{O87}.  Indeed, it is precisely this advantage of the FUV
band that allowed \GALEX\ to discover very faint star-forming
complexes well beyond the optical disk and with sensitivity better
than that provided by ground-based H$\alpha$ imaging (e.g.,
\citealt{bls11}).

The purpose of this paper is to investigate all plausible conventional
origin(s) for this offset component. In \S\ref{sec:DiffuseFUVEmission}
we review the measurements of diffuse high-latitude FUV emission,
followed by  a summary of the physics of hydrogen two-photon decay
(\S\ref{sec:TwoPhotonDecay}).  We investigate contribution to the
FUV band from line emission from the Galactic HIM, two-photon
emission from the Galactic WIM and low-velocity shocks
(\S\ref{sec:TheGalaxy}) and from fluorescence of atomic hydrogen
by Ly$\beta$ (\S\ref{sec:Fluorescence_Solar}) in the Solar System:
the Interplanetary Medium\footnote{The term IPM has two different
connotations in astronomy. In radio astronomy, IPM refers to the
solar wind that pervades the solar system out to the heliopause.
In planetary studies, the IPM stands for the very local interstellar
cloud into which the Solar system is moving.}  (\S\ref{sec:IPM}),
the Earth's atmosphere (``thermosphere";  \S\ref{sec:Thermosphere})
and the exosphere (\S\ref{sec:Exosphere}). In  \S\ref{sec:SummingUp}
we tally the contributions to the diffuse FUV background and conclude
that about two thirds (and, within uncertainties, possibly all) of
the offset component can be accounted for by these contributions.
We conclude in \S\ref{sec:ConcludingThoughts}.

\section{The Diffuse FUV Emission}
 \label{sec:DiffuseFUVEmission}
 
The ``Interstellar Radiation Field" (ISRF) is  composed of both
resolved (bright) stars and diffuse emission (e.g., Chapter 12,
\citealt{D11}).  Only the latter component is of interest to this
paper.  There have been extensive reviews of the diffuse FUV emission
(e.g. see \citealt{pmb80,H91,hmo+15}).  Here we focus on the
measurements of the diffuse FUV by \GALEX.  \GALEX\ was in low-earth
orbit (LEO) with an orbital period of 99\,minutes.  To minimize
airglow, observations were restricted to narrow periods lasting
only 25--30\,minutes during Earth eclipse \citep{mcb+07}.  The good
angular resolution of \GALEX\ allowed for masking out point sources
and galaxies.

We refer to two major studies of diffuse FUV carried out with \GALEX:
one led by Jayant Murthy and
 {associates}
(e.g., \citealt{mhs10,hmo+15,amr+18,amr+19}) and the other, the PhD
thesis of Erika Hamden \citep{hss13}.  The conclusions from these
studies are largely concordant and briefly are as follows:
 \begin{enumerate}
  \item At mid to high Galactic latitudes,
the FUV intensity scales linearly with the  IRAS $100\,\mu$m intensity
up to an intensity of $8\,{\rm MJy\,sr^{-1}}$ (e.g., \citealt{sek+11}).
This is attributed to reflection of stellar FUV photons from hot
stars (primarily located in the Galactic plane) by interstellar
dust. The same dust particles, heated by stellar light (including
the FUV), radiate at long wavelengths and produce the IRAS 100\,$\mu$m
band, hence the correlation.The FUV emission saturates at higher
values of 100\,$\mu$m emission presumably because the absorption
of the FUV radiation by the clouds outweighs over reflection.
 \item The FUV intensity increases in the direction of
interstellar clouds which harbor molecular hydrogen. This is
attributed to fluorescence by photons in the Lyman band (912--1216\,\AA)
and FUV photons (see \citealt{mhb90,jsm+17}).
 \item At high Galactic latitudes, diffuse FUV background is present,
even in directions towards dust-free regions.  
 {\citet{amr+19} estimate the EBL contribution to be
 96--131\,CU and partition it as follows: 60--81\,CU from other
 galaxies, 16--30\,CU from QSOs and $<20\,$CU from the IGM (and
 attribute this estimate to \citealt{mhb91}).}

\cite{hss13} conclude ``{\it There is a $\sim 300$\,CU FUV isotropic
offset which is likely due to a combination of air glow (probably
the dominant contributor), a small extragalactic background component
including continuum light from unresolved galaxies, and/or a Galactic
component not traced by other indicators}". 

\cite{amr+18} state ``{\it We find offsets of 230–290 photon
units\footnote{The ``photon unit" discussed here is the same as
CU.}  in the far-UV... Of this, approximately 120 photon units can
be ascribed to dust-scattered light and another 110 photon units
to extragalactic radiation. The remaining radiation is, as yet,
unidentified and amounts to 120–180 photon units in the far-UV.}"
\end{enumerate}

We end this section by noting that, at times in the literature, the
diffuse component  under discussion is referred to as ``isotropic/offset".
We do not consider the isotropic nature to be firmly established
since the primary observation is principally from high latitude
regions. The term ``offset" accurately describes the method by which
the strength of the component has been measured. So we will, for
short hand, refer to the component under investigation as the
``offset" component.  

\begin{figure}[hbtp]    %TwoPhotonSpectrum.m
 \plotone{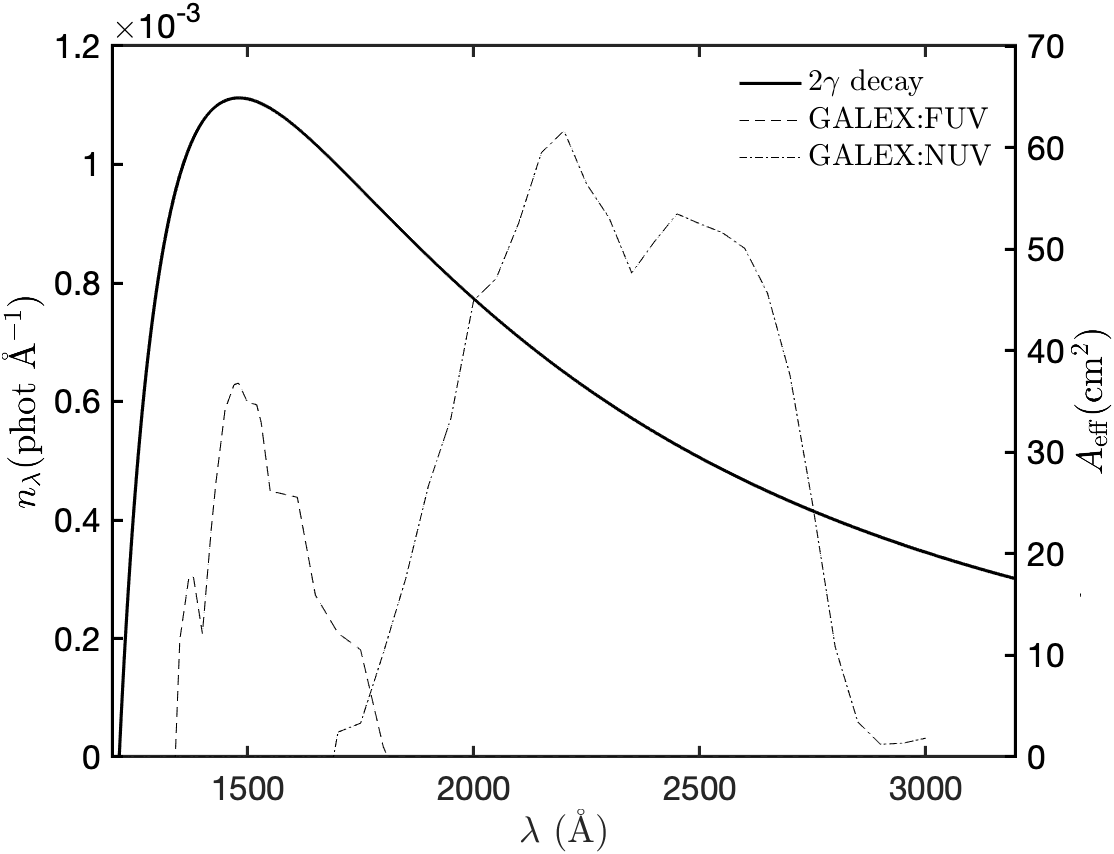}
  \caption{\small (Left $y$ axis): The photon spectrum of a single
  two-photon decay, $n_\lambda$  (solid line)  as a function of the
  wavelength, $\lambda$. Since each decay leads to emission of two
  photons, $\int n_\lambda d\lambda=2$.  (Right $y$ axis): Effective
  area of \GALEX\ FUV and NUV bands as a function of $\lambda$
  (dot-dash and dashed lines).  The details of the fitting formula
  used to generate the two-photon decay spectrum and effective area
  of \GALEX\ can be found in \S\ref{sec:Two-PhotonContinuum}.}
 \label{fig:TwoPhoton_Galex}
\end{figure}

\section{Two-Photon Decay}
 \label{sec:TwoPhotonDecay}

An H atom in a 2s level, undisturbed by a collision, will decay by
emitting two photons over a timescale of $A_{2\gamma}^{-1}\approx
(8.2\,{\rm s}^{-1})^{-1}$; see Figure~\ref{fig:TwoPhoton_Galex}.
The sum of the energies of the two photons is equal to that of
Ly$\alpha$ or 10.2\,eV.    The factors to convert the two-photon
spectrum to FUV and NUV counting rates are  central to this paper
and a full discussion can be found in \S\ref{sec:Two-PhotonContinuum}.
The key result is the following: $10^6$\,two-photon decays\,${\rm
cm^{-2}\,s^{-1}}$ results in 85.1\,CU in the FUV  channel and
18.4\,CU in the NUV channel (Table~\ref{tab:2photon_GALEX}).

\section{The Galaxy}
 \label{sec:TheGalaxy}

There are three major Galactic sources of diffuse FUV radiation:
two-photon continuum from the WIM (recombination) and shocked gas
(collisional excitation and recombination) and line emission from
the HIM.

\subsection{The Warm Ionized Medium}
 \label{sec:WarmIonizedMedium}

\citet{djb82} investigated two-photon emission from the WIM and
concluded that it could not account for the FUV background.
\citet{mhb91} and \citet{R92} and came to similar conclusions.
Rather than compute the expected two-photon emission from the
physical parameters of the WIM (as was done in the past papers) we
prefer the simpler approach of  estimating the two-photon brightness
directly from observations of H$\alpha$. At the temperature of the
WIM (8,000\,K), assuming case B, we expect that for every two-photon
decay there are 1.47 H$\alpha$ photons.

\begin{figure}[htbp]    %WHAM_Intensity.m
 \plotone{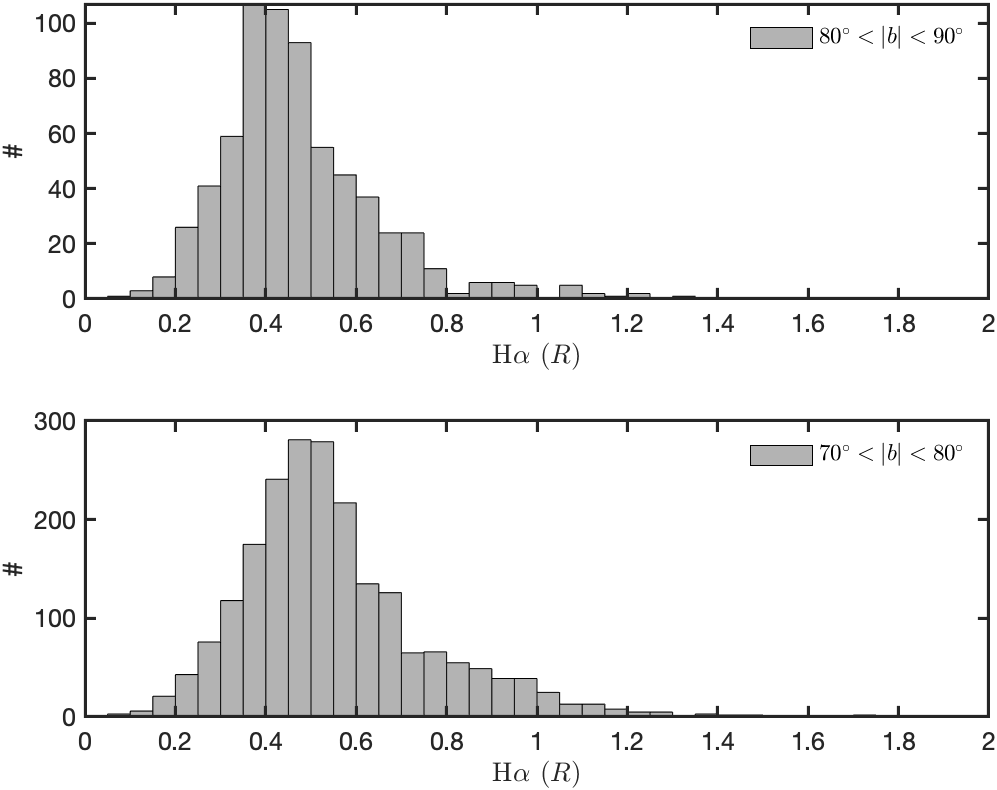}
  \caption{\small Histogram of the WHAM H$\alpha$ brightness towards
  the Galactic caps.  The vertical-axis refers to the number of
  WHAM beams.  The size of the WHAM beam is about a degree. Data
  obtained from WHAM  \citep{hrt+03,hrm+10}..  }
 \label{fig:WHAM_Intensity}
\end{figure}

In Figure~\ref{fig:WHAM_Intensity} we display the histogram of
Galactic H$\alpha$ surface brightness as recorded in the  Wisconsin
H$\alpha$ Mapper Sky Survey (WHAM-SS or WHAM for short;
\citealt{hrt+03,hrm+10}).  From these figures we conclude that the
Galactic H$\alpha$ emission towards the Galactic polar cap is
approximately 0.5\,R.  We subtract 20\% to account for contribution
resulting from scattering of H$\alpha$ emission by cirrus clouds
\citep{wgb+10,dd11}.  The expected two-photon decay rate is then
$0.27\,R$ which corresponds to 23\,CU in the FUV band,  
consistent with the estimate of 18--36\,CU \citep{sek+11}.

\subsection{The Hot Ionized Medium}
 \label{sec:TheHotIonizedMedium}

The three dominant metals  have strong resonance lines (2s$\rightarrow$2p)
in the UV,  which conveniently probe a range of temperatures
\citep{sd93}: CIV ($\lambda\lambda$\,1548.2, 1550.8\,\AA; $1\times
10^5$\,K), NV ($\lambda\lambda$\,1242.8, 1238.8\,\AA; $2\times
10^5$\,K) and OVI ($\lambda\lambda$\,1037.6, 1031.9\,\AA; $3\times
10^5$\,K). FUSE observations of OVI   established the presence of
widespread HIM in the halo of the Galaxy \citep{ssj+00} and in the
disk of the Galaxy \citep{bjt+08}.

 	%CHIANTI_plasma_plot.m

FIMS/SPEAR aboard South Korea's STSAT-1 satellite undertook all sky
spectral-imaging in the wavelength range 900--1150\,\AA\ and
1350--1750\,\AA\ \citep{sek+11,jsm+17,jsm+19}.  The authors fit
high latitude data to a collisional-ionization equilibrium (CIE)
model and find a plasma temperature of $2\times 10^5$\,K and a
vertical emission measure of $\int n_e(z)^2dz \approx 0.01\,{\rm
cm^{-6}\,pc}$.  For these parameters, the free-free emission amounts
to $<1\,$CU in the FUV band.  The most prominent line in the spectrum
towards the Galactic poles is the CIV doublet with a strength of
4,600\,LU. Separately, \citet{mhb91} report a similar detection
towards the Lockman hole -- a region with a very low (lowest?) H~I
column density. We will thus assume that 4,600\,LU is not reflected
light but is genuine emission from high-latitude HIM.

In order to determine the total emission from the HIM we need to
account for fainter lines and continuum (free-bound, primarily).
To this end,  we used CHIANTI \citep{dlm+97,ddy+21} and obtained
the CIE spectrum of a solar-abundance plasma at $T=[1,2]\times
10^5\,$K.  We integrated this model with the effective area of the
\GALEX\ FUV detector (see Figure~\ref{fig:TwoPhoton_Galex}) and
found that the CIV emission is $[40\%, 80\%]$ of the emission
integrated over the FUV band.  Taking the average we find the  line
emission from the HIM contributes $1.875\times 4600=8625\,$LU to
the FUV band.  The effective width of the \GALEX\ FUV band is
255\,\AA\ (Table~\ref{tab:2photon_GALEX}).  Thus, the HIM contribution
to the FUV background is 34\,CU.

\subsection{Low-velocity Shocks}
 \label{sec:LowVelocityShocks}

Shocks are a central topic in the study of the ISM. The post-shocked
gas cools via line emission, free-bound and free-free emission.
Spectral libraries of  cooling gas have been published with
increasing sophistication; see \citet{R79}, \citet{sm79} and
\citet{sd93}.  Shocks with low-velocities, $v_s\lesssim 100\,{\rm
km\,s^{-1}}$, cool via lines of H~I (in decreasing order, Ly$\alpha$,
two-photon continuum, H$\alpha$).  In fact, one of the earliest and
unambiguous detection of two-photon emission was from a Herbig-Haro
object with a 100\,km\,s$^{-1}$ shock (\citealt{bss82}; see also
\citealt{dbs82}).

With some interest I note that two-photon emission was invoked in
a recent investigation of the fascinating  30$^\circ$-long H$\alpha$
arc in Ursa Major \citep{mb01,bba+20}.  Separately,  \citet{fdw+21},
using \GALEX\ FUV observations along with H$\alpha$ and radio
imaging, uncovered three very large supernova remnants. In both
cases, the underlying cause of two-photon emission is low-velocity
shocks, $v\lesssim 100\,{\rm km\,s^{-1}}$.

All high velocity shocks eventually cascade to low-velocity shocks.
In fact, the classical explanation for the observed $\approx 10\,{\rm
km\,s^{-1}}$ velocity dispersion of the CNM is the stirring of the
ISM by supernovae (\S12.2c, \citealt{S78,ko15}).  The mass of the
Galactic H~I gas is $M_{\rm HI}=3\times 10^9\,M_\odot$ (Chapter 1
of \citealt{D11}).  The Galactic supernova (SN) rate is 
estimated to be one per century or $\dot{E}_{\rm SN}=3\times
10^{41}\,{\rm erg\,s^{-1}}$, assuming $E_0=10^{51}\,$erg per SN.
Let $\eta$ be the conversion efficiency of SN energy that goes into
the stirring of the atomic phases and let $\sigma_{\rm HI}$ be the
effective rms velocity of CNM+WNM. Then the mean time between
successive stirrings is
 \begin{equation}
	\tau_{\rm HI} = \frac{\sfrac{1}{2}M_{\rm HI}\sigma_{\rm
	HI}^2}{\eta\dot{E}_{\rm SN}}\approx
	10^{13}\eta^{-1}\sigma_6^2\,{\rm s}, \nonumber
 \end{equation}
where $\sigma_{\rm HI}=1\times 10^6\sigma_6\,{\rm cm\,s^{-1}}$.
Locally, the vertical column density towards the Galactic poles is
$N_H\approx 2\times 10^{20}\, {\rm atom\,cm^{-2}}$ \citep{lg91}.
Thus, the rate of successive SN impacts in a vertical column is
 \begin{equation}
  \mathcal{F}_{\rm SN}=N_H/\tau_{\rm HI}= 2\times
  10^7\eta\sigma_6^{-2}\,{\rm atom\,cm^{-2}\,s^{-1}}.  \label{eq:F_SN}
 \end{equation}

In order to convert $\mathcal{F}_{\rm SN}$ to two-photon decays we
make the following (conservative) assumptions: (1) we restrict to
shocks with initial  velocity, $v\gtrsim 100\,{\rm km\,s^{-1}}$.
Such shocks will pre-ionize the ambient gas \citep{R79,sm79}. (2)
In the post-shocked region each H ion undergoes only one recombination,
(2) about a third of the  recombinations lead to two-photon emissions
and (3) following recombination,  no H atom is collisionally excited
to the 2s state or 3p state.  With these assumptions, the columnar
rate of two-photon decay is $\mathcal{R}\approx
\sfrac{1}{3}\mathcal{F}_{\rm SN}$.

\citet{mo77}, in the framework of the 3-phase model for the ISM,
found $\eta=0.05$.  \citet{ko15} carry out a detailed simulation
and find that each SN (with $E_0=10^{51}\,$erg) has, at the onset
of the formation of radiative shells (``snowplow'' phase), a velocity,
$v_{\rm sf}=200\,{\rm km\,s^{-1}}$ and a momentum of
$p_{\rm sf}=2\times 10^5 n_0^{-0.15}\,{M_\odot\rm\, km\,s^{-1}}$
where $n_0$ is the  particle density of the medium into which the
SN exploded.  If we assume that all this momentum results in stirring
of clouds then $\eta=1/2(p_{\rm sf}/\sigma_{\rm HI})\sigma_{\rm
HI}^2/E_0 \approx 2\sigma_6\%$.  
We adopt mean of the
two estimates $\eta=0.035$ and find
$\mathcal{R}=0.23\times10^6\sigma_6^{-2}\, {\rm decay\,cm^{-2}\,s^{-1}}$.
This corresponds to an FUV brightness of $20\,$CU.

\section{Fluorescence of Ly$\beta$ photons}
 \label{sec:Fluorescence_Solar}
 
\begin{figure}[htbp]    %HI_partialGrotrian_II.m
 \centering \includegraphics[width=2in]{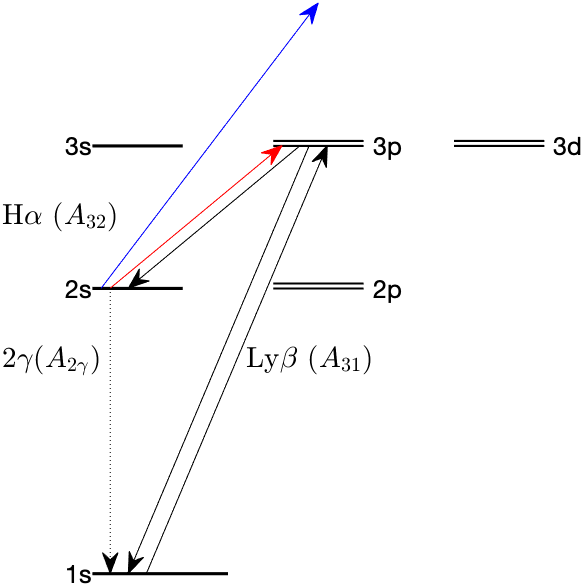}
  \caption{\small Partial Grotrian diagram for H~I (not to scale).
  An H atom excited by Ly$\beta$ to 3p can de-excite to ground state
  by emitting Ly$\beta$ (probability of $\approx 7/8$) or de-excite
  to the $2{\rm s}\ ^2S_{\sfrac{1}{2}}$ level by emitting H$\alpha$.
  An atom in the 2s state can be further  excited by Balmer photons
  (e.g. H$\alpha$, red arrow) or bound-free
  process (blue arrow); see \S\ref{sec:Excitation_2s_Solar} for
  discussion.}
 \label{fig:Ly_beta_Grotrian}
\end{figure}

In the Solar system, two-photon emission is a result of fluorescence
of solar Ly$\beta$ absorbed by H atoms.  There is a rich literature
of both theory and observations related to  diffuse Ly$\alpha$ in
the Solar system and less so on Ly$\beta$.  Following excitation
by a Ly$\beta$ photon to one of the 3p levels the atom can relax
to the ground state by  emitting a Ly$\beta$ photon (we denote the
corresponding A-coefficient by $A_{31}$) or de-excite to the $2{\rm
s}\,^2S_{\sfrac{1}{2}}$ state by emitting an H$\alpha$ photon
($A_{32}$); see Figure~\ref{fig:Ly_beta_Grotrian}. The branching
ratio to emit Ly$\beta$ is $B=A_{31}/(A_{31}+A_{32})=0.88$. We let
$\eta_{H\alpha}\equiv A_{32}/A_{31}\approx 1/7.4$.  For higher order
Lyman series photons the probability to reach the $2{\rm
s}\,^2S_{\sfrac{1}{2}}$ level is smaller.  Furthermore, an inspection
of the solar spectrum shows that the intensity of the higher lines
is smaller than that of the Ly$\beta$ line. As a result, we simplify
by restricting our analysis to  only Ly$\beta$ excitations.

\begin{figure}[htbp] %SOHO_Lymanbeta_profile.m
 \plotone{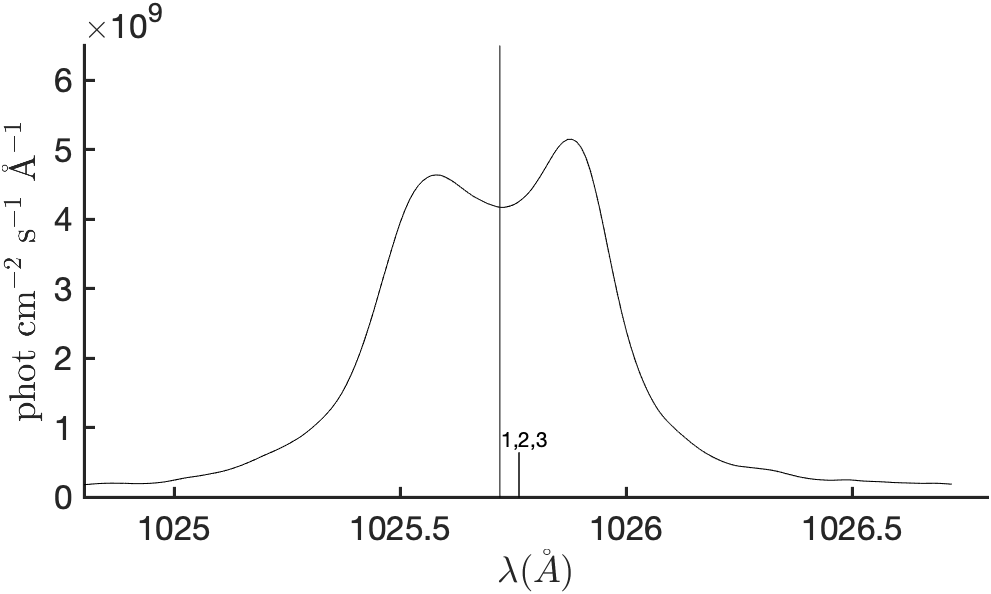}
  \caption{\small The profile of solar Ly$\beta$ obtained from the
  SOHO missoin \citep{lvc+15}. Notice the low continuum level
  (see Figure~2
  of \citealt{dla+05} for an annotated EUV/FUV solar spectrum).
  The vertical line marks the rest wavelength of Ly$\beta$.  
   {
  The three vertical ticks labeled ``1,2,3" mark the wavelengths of 
  three resonance lines of O~I which have Bowen
  fluorescence with Ly$\beta$ (see  \S\ref{sec:OI_BowenFluorescence}).}
  }
 \label{fig:Solar_Lyman_beta}
\end{figure}

The solar Ly$\beta$ line has a velocity profile with two horns
separated by about 0.33\,\AA\ and a full-width-at-zero of 1\,\AA\
(Figure~\ref{fig:Solar_Lyman_beta}).  In other words, the velocity
width of Ly$\beta$ is quite narrow, $\pm 50\,{\rm km\,s^{-1}}$. 
The geo-coronal H atoms, with respect to the Sun, have
essentially zero radial velocity and a thermal velocity full-width
at half-maximum (FWHM)  of 7\,km\,s$^{-1}$.  The H atoms in the IPM
have a mean velocity of $-24\,{\rm km\,s^{-1}}$ with respect to the
Sun. The thermal velocity spread of those H atoms is $18\,{\rm
km\,s^{-1}}$. Thus, H atoms in the geo-coronal and IPM can readily
absorb solar Ly$\beta$ photons. 

A linear relation exists between the intensity at the valley center
and the integrated Ly$\beta$ emission.  This relation allows solar
astronomers to conveniently infer the zero velocity (central)
Ly$\beta$ intensity from integrated Ly$\beta$ observations. 
Over
the period 1996--2009 (solar cycle 23), the central photon flux
density at 1\,AU varied from $4\times 10^{10}\,{\rm
phot\,cm^{-2}\,s^{-1}\,nm^{-1}}$ to $10\times 10^{10}\,{\rm
phot\,cm^{-2}\,s^{-1}\,nm^{-1}}$ \citep{lvc+15}.  We adopt the
minimum flux for our fiducial value. We also switch to the
traditional \AA\ unit  for the  differential wavelength 
and thus
$F_\lambda(\beta,v=0)=4\times 10^{9}\,{\rm phot\,cm^{-2}\,s^{-1}}\,$\AA; 
here, $v$ is the radial velocity between the
absorbing H atoms and the mean velocity of the Sun.  
The spectral frequency flux density is 
 \begin{equation}
   F_\nu(\beta,v=0)=1.4\times 10^{-3}\,{\rm
   phot\,cm^{-2}\,s^{-1}\,Hz^{-1}}.
     \label{eq:F_solar_beta}
  \end{equation}
Going forward, we use the short hand $F_\nu(\beta,v=0)=F_\nu(0)$
and so on.

The frequency-dependent absorption cross-section is $ \sigma_{lu}(\nu)
=(\pi e^2)/(m_ec) f_{lu}\phi_\nu$ where  $l$ ($u$) stands for lower
(upper) levels, $f_{lu}$ is the oscillator strength and  $\phi_\nu$
is the probability distribution as a function of frequency for the
absorption process, $\int\phi_\nu d\nu=1$.  The Ly$\beta$ column
densities in the IPM and exosphere are modest and 
so we can entirely ignore the ``damping" or Lorentzian wings.    If,
as it happens to be the case, the solar Ly$\beta$ photon intensity is constant over
the thermal frequency spread of Ly$\beta$ absorption frequency of 
terrestrial H atoms then $F_\nu=F_\nu(0)$ and the
rate of excitation per atom is
 \begin{eqnarray}
	\mathcal{R} &=& \int F_\nu\sigma_{lu}(\nu) d\nu =
	F_\nu(0)\frac{\pi e^2}{m_ec}f_{lu}\ {\rm  atom^{-1}\,s^{-1}}.
	\label{eq:R_excitation}
 \end{eqnarray}

Consider a  medium that is optically thin and has sufficiently low
particle density so that the probability of an H atom colliding
with another particle, particularly a proton, over a duration of
$A_{2\gamma}^{-1}$, is negligible.  Using Equations~\ref{eq:F_solar_beta}
and Equation~\ref{eq:R_excitation} and noting that the oscillator
strength is $f_{lu}=0.079$ we find the total rate of excitation to
the 3p state is
 \begin{equation}
  R_\beta = 2.94\times 10^{-6}d_{\rm AU}^{-2}\,{\rm atom^{-1}\,s^{-1}}.
  \label{eq:R_beta}
 \end{equation}
The corresponding Ly$\beta$  photon surface brightness is
$L_\beta=\mathcal{R}_\beta N_H$ where $N_H$ is the hydrogen column
density, along the line of sight.

The ratio of the solar Ly$\alpha$ to Ly$\beta$ intensity (with
intensity expressed in energy units and not photons) ranges from
50 (solar maximum) to 80 (solar minimum); see  \citep{lvc+12}.  Thus
the photon flux ratio of Ly$\beta$ to that of Ly$\alpha$, hereafter
$\eta_\odot$, ranges from 1/95 to 1/60.  We adopt, the harmonic
mean $\eta_\odot=1/73$.  The ratio of $\lambda^2f_{lu}$ for the two
lines is 7.4.  Thus, for optically thin medium, $L_\beta\approx
L_\alpha/540$.  The surface brightness in H$\alpha$, $L_{H\alpha}=
\eta_{H\alpha}L_\beta\approx L_\alpha/4000$. 
 If, the
medium has low-density (see \S\ref{sec:CollisionalMixing}), which
is the case for IPM and exosphere, then every H$\alpha$ emission
is accompanied by two-photon emission.  Thus, $L_{2\gamma}=L_{H\alpha}$,
where both quantities are expressed in photon units.

However, in the inner solar system, there is a copious flux of solar
Balmer photons (H$\alpha$, H$\beta$...). 
As can be seen from Figure~\ref{fig:Ly_beta_Grotrian} an H atom in the
2s level can be further excited to higher levels by absorbing a Balmer
photon (``solar pumping"). In \S\ref{sec:Excitation_2s_Solar} we
show that this process results in a pumping rate, $A_p\approx
2.2d_{\rm AU}^{-2}\,{\rm s}^{-1}$ where $d_{\rm AU}$ is the distance
of the hydrogen atom from the Sun in units of AU.  
An H-atom,  upon absorption of Ly$\beta$, now has several outcomes:
decay to ground state by emitting Ly$\beta$, decay to 2s followed by
pumping back to 3p and then followed by decay to ground state by
emitting Ly$\beta$ or photo-ionized or excited to 4p state or finally
decaying to ground-state via two-photon continuum.
So, instead of having a single branching ratio we now have several.
The bottom line is that 
solar 
pumping  reduces the two-photon emission probability by
$p_{2\gamma}=A_{2\gamma}/(A_{2\gamma}+A_p)$.  
Incidentally, this means in the inner solar system Ly$\beta$ fluorescence
will have reduced two-photon emission but brighter H$\alpha$ emission 
(cf.\ \citealt{ssr84}).  

\section{Interplanetary Medium}
 \label{sec:IPM}

The entire Solar system is moving into a local interstellar cloud
at a velocity of about 25 km\,s$^{-1}$. From high-resolution
spectroscopic UV studies the following properties of the interstellar
cloud have been deduced: neutral H density of 0.1\,cm$^{-3}$,
temperature of 7000\,K and 50\% ionization fraction \citep{frs11,gj17}.
These properties seem to have  physical attributes similar to the
WNM, albeit with a high fractional ionization and also a low column
density.  While astronomers refer to this cloud as the ``very local"
ISM (VLISM) planetary astronomers call this medium as the
interplanetary medium (IPM).

The Sun has a weak wind, about $10^{-14}\,M_\odot\,{\rm yr}^{-1}$.
The wind is correlated with solar activity and also has dependence
on solar latitude and longitude \citep{pb04}. At 1\,AU, the typical
properties are as follows: electron density, $n_e$  of 3--10\,${\rm
cm^{-3}}$, magnetic field strength of 10--370\,$\mu$G and temperature
of $10^5\,$K. By the time the wind reaches Earth it is supersonic,
$v_{\rm wind}\approx 500\,{\rm km\,s^{-1}}$.  

The fast solar wind comes to equilibrium with the slow moving
interstellar medium by undergoing a shock -- the so-called solar
wind termination shock (SWTS).  In the parlance of planetary
astronomers, the surface separating the  shocked solar wind and the
interstellar gas is called the heliopause (aka ``contact discontinuity"
for astronomers). Thanks to the Voyager missions we know that the
SWTS is located at 90\,AU while the heliopause is at 120\,AU.

The neutral particles of the VLISM/IPM are not affected by the shock
fronts and drift into solar system.
The incoming H atoms scatter solar Ly$\alpha$ and thereby creating
a Ly$\alpha$ haze over the sky, first detected by  
OGO-5  (see \citealt{bb71,tk71}).   The direction of the
wind has been precisely determined (e.g., \citealt{pbl+13,gj17}).
The ``downwind" direction in ecliptic longitude ($\lambda)$ and
latitude ($\beta$) coordinates is $79^\circ$ and $-5^\circ$ which
corresponds to Galactic coordinates $l=185^\circ$ and $b=-12^\circ$.
The upwind direction, $l=5^\circ$ and $b=12^\circ$, corresponds to
a direction just ``above" Scorpius constellation but ``below"
Ophiuchus.  The H atoms are photo-ionized ($\mathcal{R}_{\rm
pi}=0.8$--$0.17\times 10^{-7}d_{\rm AU}^{-2}\,{\rm s^{-1}}$) or
undergo charge exchange\footnote{Following charge exchange the H
atom acquires the erstwhile large velocity of solar wind protons
and can thus no longer absorb solar Ly$\beta$ flux with its narrow
velocity width.} ($\mathcal{R}_{\rm ce}\approx 4\times 10^{-7}d_{\rm
AU}^{-2}\,{\rm s^{-1}}$)  as they approach the sun.  The Solar Wind
ANistropies (SWAN) instrument aboard the SOlar Heliospheric Observatory
(SOHO) uses a hydrogen cell to map the Ly$\alpha$ haze (see
\S\ref{sec:SWAN-IMAGE} for summary). The SWAN sky
image\footnote{\url{http://swan.projet.latmos.ipsl.fr/}} is highly
asymmetric, bright in the upwind direction and faint in the downwind
direction.

\subsection{Measurements}

Voyager-2 was launched on 1977 August 20. One month later (heliocentric
distance\footnote{The distance to Voyager-2 was computed using a
tool provided at \url{https://omniweb.gsfc.nasa.gov/coho/helios/heli.html}}
of 1.02\,AU) the UV spectrometer was pointed to $\alpha=324^\circ$
and $\delta=-23^\circ$ and  detected Ly$\alpha$ at $722\pm 0.5\,R$,
Ly$\beta$ at $2\pm 0.16\,R$ and Helium 584\,\AA\ at $3.8\pm 0.04\,R$
\citep{ssb78}.  Long observations of several high Galactic latitude
fields were undertaken in February and March, 1981 (8.4\,AU).  These
observations detected not only Ly$\beta$ but also higher order Lyman
lines \citep{H86}.  Of interest to us were $L_\alpha=1080\,R$ and
$L_\beta=2.4\,R$.  The ratio between Ly$\alpha$ and Ly$\beta$ seems
to be consistent with that expected for optically thin medium
(\S\ref{sec:Fluorescence_Solar}).  The New Horizons mission also
carried a UV spectrometer and detected Ly$\alpha$ surface brightness
of $550\,R$ at a heliocentric distance of 7.6\,AU decreasing to
$100\,R$ at 38\,AU \citep{gps+18} -- confirming that peak Ly$\alpha$
production  takes place in 2--5\,AU (in the upwind direction).
Incidentally, the latest report from New Horizons
(now at 48\,AU) is the detection of an isotropic Ly$\alpha$ glow
with a brightness of about 43\,$R$ and  ascribed to Ly$\alpha$
leaking from the vicinity of distant hot stars \citep{gph+21}.

We now summarize this section. From our vantage point, in the
``upwind" direction of the IPM, we  see $L_\beta=2.4\,R$.  The
expected two-photon decay rate is $\eta_{H\alpha}L_\beta=0.33\,R$.
The corresponding brightness in the FUV band is about 28\,CU.

\section{The Atmosphere of Earth}
 \label{sec:Thermosphere}
 
The same solar Lyman photons that excite the IPM also excite hydrogen
atoms in the thermosphere and  the exosphere.
The key
regions of the atmosphere relevant to this section are thermosphere,
exosphere and ionosphere (see \S\ref{sec:Thermosphere_Appendix},
\S\ref{sec:TheIonosphere} and \S\ref{sec:Exosphere_Appendix}
for background summaries).  In
the thermosphere (100--600\,km) atomic hydrogen is a minority
species. Solar EUV heat up this region by cleaving molecules. At
the top of the thermosphere (``exobase") the temperature is high
enough for H and He atoms  to achieve sufficiently high
thermal velocity to escape from
Earth. The resulting slowly outflowing exosphere is detectable to
distance beyond the lunar orbit.

Ly$\beta$, due to a coincidence, can also excite atomic oxygen (aka
Bowen fluorescence; see \S\ref{sec:OI_BowenFluorescence}).  Next,
as will be explained below, two-photon production (but not H$\alpha$)
is suppressed at proton densities higher than $10^4\,{\rm cm^{-3}}$.
As a result of these two complications, our estimate of two-photon
brightness from the thermosphere and the exobase is uncertain. We
can, however, with confidence, estimate the two-photon output from
a height of 1500\,km and above. It is this estimate that is presented
in Table~\ref{tab:Inventory} under the heading ``exosphere".

\begin{figure}[htbp] 	%ShadowHeight.m
 \plotone{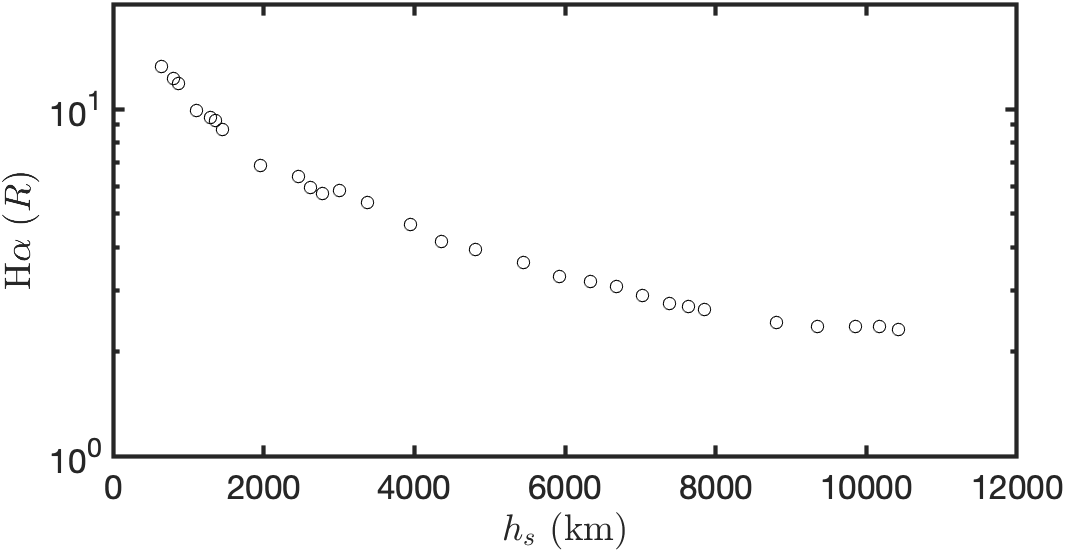}
  \caption{\small Representative zenith angle scan of geo-coronal
  H$\alpha$ undertaken with WHAM at Kitt Peak National Observatory
  at dusk shown as a function of shadow height, $h_s$.  The WHAM
  data are from \citet{nmr+08} and obtained during February 2000
  (solar maximum). }
 \label{fig:ZenithAngle_Halpha_Scan}
\end{figure}

\subsection{Ground-based observations of geocoronal H$\alpha$}

\begin{figure}[htbp]    %ShadowGeometry.m
 \plotone{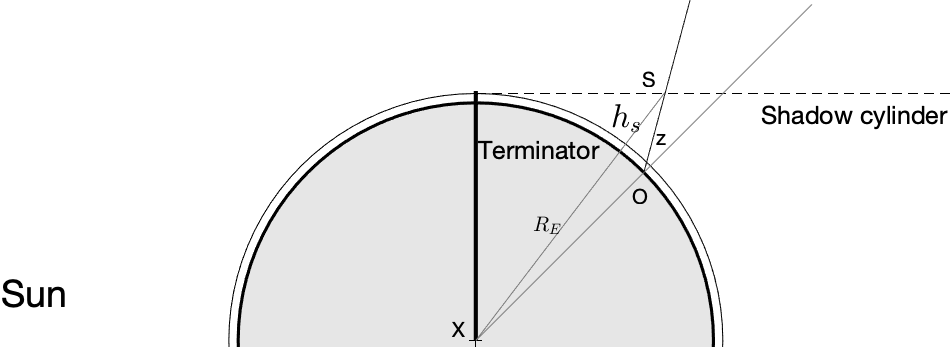}
  \caption{\small Illustration to explain the concept of shadow
  height or shadow altitude.  The thick black circle represents the
  surface of Earth (radius, $R_E$) while the  lighter circle
  represents the height at which Ly$\beta$ is absorbed (about
  100\,km; primarily by molecular oxygen).  The Sun (left) casts a
  cylindrical shadow (dashed lines) to the right.  The line-of-sight
  of the observer (``O") bearing zenith angle $z$ intersects the
  shadow cylinder at ``S".  The ``shadow" height, $h_s$, is the
  radial distance above the Earth's surface to point S, $h_s=x_s-R_E$
  where $x_s$ is the length of segment XS. }
 \label{fig:Geometry_ShadowHeight}
\end{figure}

WHAM, during dusk and dawn, undertook scans of the sky in H$\alpha$
as a function of zenith angle for aeronomical studies. These scans, 
by tradition, are displayed as  a function of ``shadow height"
(Figure~\ref{fig:Geometry_ShadowHeight}).  Next, note that within the
shadow cylinder, in the absence of multiple scattering, no H$\alpha$
emission will take place.  However, Ly$\beta$+OI and Ly$\alpha$ is
clearly seen by LEO missions looking ``upward" (see
\S\ref{sec:LowEarthMissions} and \S\ref{sec:OI_BowenFluorescence}).  Detailed modeling \citep{bhn+01}
show, in fact that, within the thermosphere (100-500\,km) scattered
Ly$\beta$ is comparable to the incident flux.  Note also that
``cascade" or contribution to H$\alpha$ due to recombinations from
higher levels is estimated to be less than 10\% \citep{bhn+01}. So,
it is reasonable to assume that H$\alpha$ is essentially due to
fluorescence of Ly$\beta$.

The H$\alpha$ brightness at large shadow heights (cf.\
Figure~\ref{fig:ZenithAngle_Halpha_Scan}) was found to be 1.5\,$R$
at solar minimum and twice that at  solar maximum \citep{nmr+08,nmr+19}.
The corresponding two-photon brightness, using the formulation
discussed in \S\ref{sec:TwoPhotonDecay}, is 85\,CU which, when
summed with other sources of two-photon emission (\S\ref{sec:TheGalaxy},
\S\ref{sec:IPM}), would exceed the level of the offset emission of
120--180\,CU.  This apparent contradiction is a result of ignoring
a key physical process that arises in dense regions and is discussed
next.

\subsection{Collisional Mixing of 2s and 2p states}
 \label{sec:CollisionalMixing}
 
Given the milli-eV energy difference, a collision can easily shift
an H atom from the 2s to one of the two 2p states after which the
atom will, in short order ($A^{-1}_{21}\approx 2\,{\rm ns}$), emit
a Ly$\alpha$ photon and thereby suppress two-photon emission.  For
neutral particle colliders we adopt a ``hard sphere" model and
set the cross-section to $\sigma=10^{-16}\,{\rm cm^2}$.  We find
that the time scale for collisions is longer than $A_{2\gamma}^{-1}$
throughout the thermosphere (see right panel of
Figure~\ref{fig:Thermosphere_Profile}).  It turns out that the
slower moving protons are more effective than electrons in effecting
these (primarily) angular-momentum changing 2s$\rightarrow$2p
transitions (\citealt{P52}; see \S\ref{sec:2s_2p_collisional_mixing}).
A reasonable night-time temperature for the thermosphere is $\approx
1,000$\,K (\S\ref{sec:Exosphere_Appendix}).  Let $q_{e+p}$ be the
sum of the collisional coefficients for electrons and
protons.\footnote{Ions such as O$^+$ are even more effective because
of their slower velocities. However, by 1,000\,km heigh the dominant
species is H and not O; see Figure~\ref{fig:Thermosphere_Profile}.}
Matching the inverse mean time between collisions $n_pq_{e,p}$ to
$A_{2\gamma}$ yields the critical density $n_{\rm crit}=A_{2\gamma}/q_{e,p}
\approx 10^4\,{\rm cm^{-3}}$ where $n_p$ is the mean proton density.
The production of two-photons becomes inefficient by
$(1+n_p/n_{\rm crit})$. From Figure~\ref{fig:Electron_Profile} we
conclude that two-photon emission is suppressed below 1,200\,km.
However, note that this collisional mixing does not affect the
production of H$\alpha$ photons.

\subsection{Observations of Ly$\beta$ in the Thermosphere}
 \label{sec:LowEarthMissions} 

Two missions, 
 {
both in low-earth (height of about 600\,km) } orbit, Space Test
Program 78-1 (STP\,78-1; aka ``Solwind") and EURD aboard the Spanish
Minisat-01 mission \citep{mgt+98}, measured Ly$\beta$ brightness.
{Details of  the missions  are summarized in
\S\ref{sec:LEO-satellites}.}

At the beginning of this section we alerted the reader of Bowen
fluorescence of Ly$\beta$ by O~I. In detail,  three lines of O~I
(hereafter, the ``trio") lie only about $+{\rm 8\,km\,s^{-1}}$ of
the rest wavelength of Ly$\beta$ (see Figure~\ref{fig:Oxygen_Bowen})
and are thus readily excited by solar Ly$\beta$ photons (see
Figure~\ref{fig:Solar_Lyman_beta}).  Each such excitation results
in re-emission of the incident photon (71\%) which will be
indistinguishable from Ly$\beta$ or fluorescence via emission of
the O~I 1304\,\AA\ triplet (see Figure~\ref{fig:Oxygen_Bowen}).
Next, the sum of the oscillator strengths for the trio is $f_{\rm
trio}=0.0201$ which can be compared with $f_\beta=0.0791$, the
oscillator strength for Ly$\beta$.  However, as can be seen from
Figure~\ref{fig:Thermosphere_Profile}, O~I dominates over H~I up
until 800\,km.

 {The measurements are summarized in Table~\ref{tab:STP_78}.}
The observed upward Ly$\beta$+OI brightness sets an upper limit to
the surface brightness of Ly$\beta$.  The brightness of the resonant
$\lambda 1304\,$\AA\ (note that terrestrial O~I atoms also scatter
solar 1304\,\AA\  photons) means that a good fraction of the observed
Ly$\beta$+OI must be due to O~I.  On the other hand, for optically
thin conditions, the ratio of Ly$\alpha$ to Ly$\beta$ is 400 which
should be compared with 540 (\S\ref{sec:Fluorescence_Solar}). If
we assume optically thin conditions for Ly$\beta$ then we would
expect 3/4 of the upward brightness is due to Ly$\beta$.

\begin{deluxetable}{lrr}[htb]
\tablecaption{STP\,78-1 \&\ EURD observations of night glow}
\tablewidth{0pt}
\tablehead{
\colhead{line} & 
\colhead{``Up'' ($R$)} &
\colhead{``Down" ($R$)}
}
\startdata
Ly$\beta$+OI   & $8.76\pm 0.3$  & $2.3\pm 0.23$ \\
Ly$\alpha$       & $3533\pm 5.8$ & $1712\pm 4.7$ \\
OI\,1304           & $7.1\pm 0.4$    &  $53.8\pm 1.3$ \\
OI\,1356           & $<2$                 & $52.2\pm 0.8$ \\
\enddata
\tablecomments{
Restricting to absolute zenith angle of $<70^\circ$, EURD
detected, at night time, Ly$\beta$  at a level of $6.4\,R$ \citep{lmg+01}.
}
 \label{tab:STP_78}
\end{deluxetable}

The downward Ly$\beta$+OI surface brightness of 2.3\,$R$ and the
(upward) ground-based geo-coronal H$\alpha$ emission of $1.5\,R$
(solar minimum) or $3\,R$ (solar maximum) can only be reconciled
by ascribing most of the ground-based (integrated) H$\alpha$ emission
to an altitude above the orbital height of Solwind.  However,
\citet{bhn+01} argue that bulk of this H$\alpha$ arises from atomic
hydrogen between 100 and 500\,km. Their model hydrogen column density
over this range of height  of $6\times 10^{13}\,{\rm cm^{-2}}$ is
five times larger than that provided by the standard MSIS (Mass
Spectrometer and Incoherent Scatter radar) model (see
Figure~\ref{fig:Thermosphere_Profile}).  Overall, the situation is
unsatisfactory.  There is a compelling need for space-based
measurements in H$\alpha$ (\S\ref{sec:ConcludingThoughts}).  

\begin{figure}[htbp]
 \centering
  \plotone{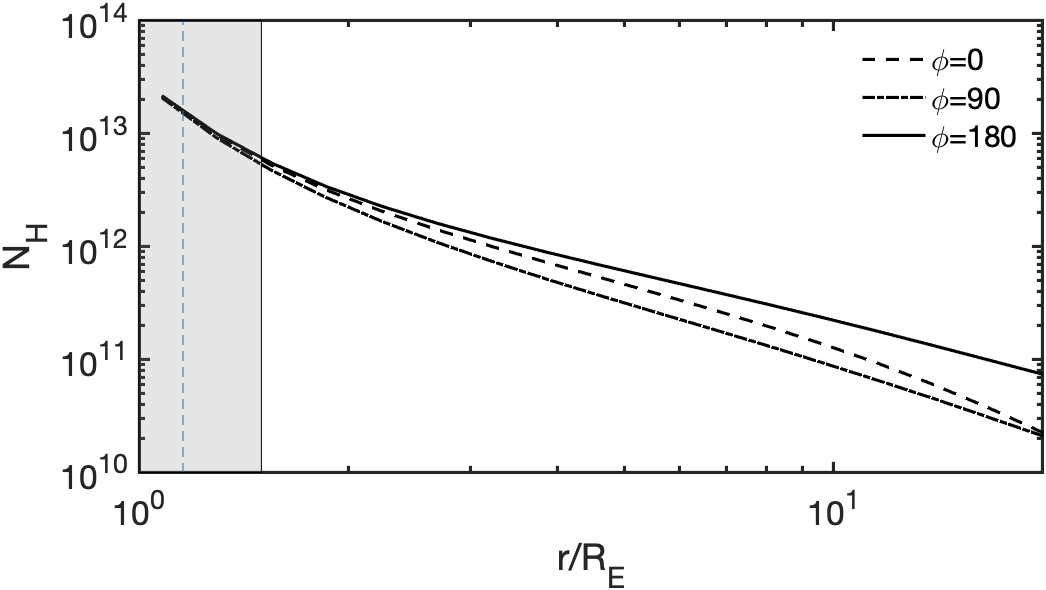}  %exoprofile_SWAN.m
   \caption{\small {\bf Top:} The column density profile of H atoms
   as derived from {\it SWAN-SOHO} measurements \citep{bbq+19}.
   Here, $r$  is the distance from center of Earth and $R_E$ is the
   radius of Earth. The dashed vertical line marks height of
   1,000\,km.  Owing to the angular resolution of {\it SWAN} the
   model is not reliable for radius less than 1.5\,$R_E$ (shaded
   region).  $\phi$ is the angle between line-of-sight and the
   direction to the Sun.  Thus the night-time radial profile is
   described by $\phi=180^\circ$ model whereas $\phi=0^\circ$ applies
   to the model at noon time.  The exosphere has a larger radial
   extent on the night side, compared to the day side (``geotail").
   The model data were supplied by I.\ Baliukin.}
 \label{fig:SWAN_Model}
\end{figure}

\section{The Exosphere}
 \label{sec:Exosphere}

We use data from two missions: the afore-mentioned SWAN on SOHO and
the IMAGE missions, the summaries for which can be found in
\S\ref{sec:SWAN-IMAGE}.  As can be seen from Figure~\ref{fig:SWAN_Model}
the hydrogen column density above $1,000\,$km is approximately
$10^{13}\,{\rm cm^{-2}}$.  The corresponding central optical depth
in Lyman-$\beta$ is given by
 \begin{equation}
  \tau_0=0.32\Big(\frac{N_l}{10^{14}\,{\rm cm^{-2}}}\Big)\Big(\frac{5\,{\rm
  km\,s^{-1}}}{b}\Big)
 \end{equation}
where $b=\sqrt{2}\sigma_v$  and $\sigma_v$ is the Gaussian velocity
width.  Thus, $\tau_0$ is less than unity and so the two-photon
brightness is $\eta_{H\alpha}L_\beta$.

\begin{figure*}[hbtp]	%Matlab: Exosphere_Geometry.m
				%Matlab: exoray.m
 \centering
  \includegraphics[height=1.5in]{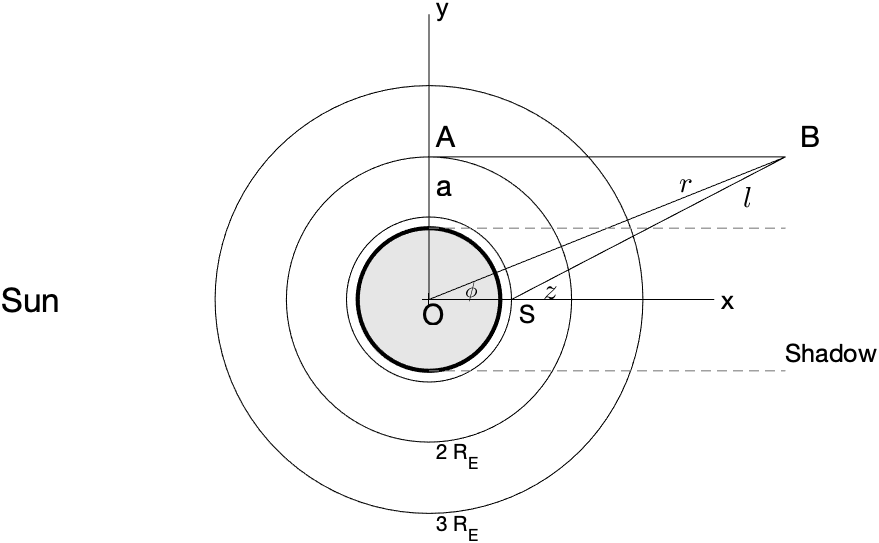}\qquad\qquad\qquad
    \includegraphics[height=1.3in]{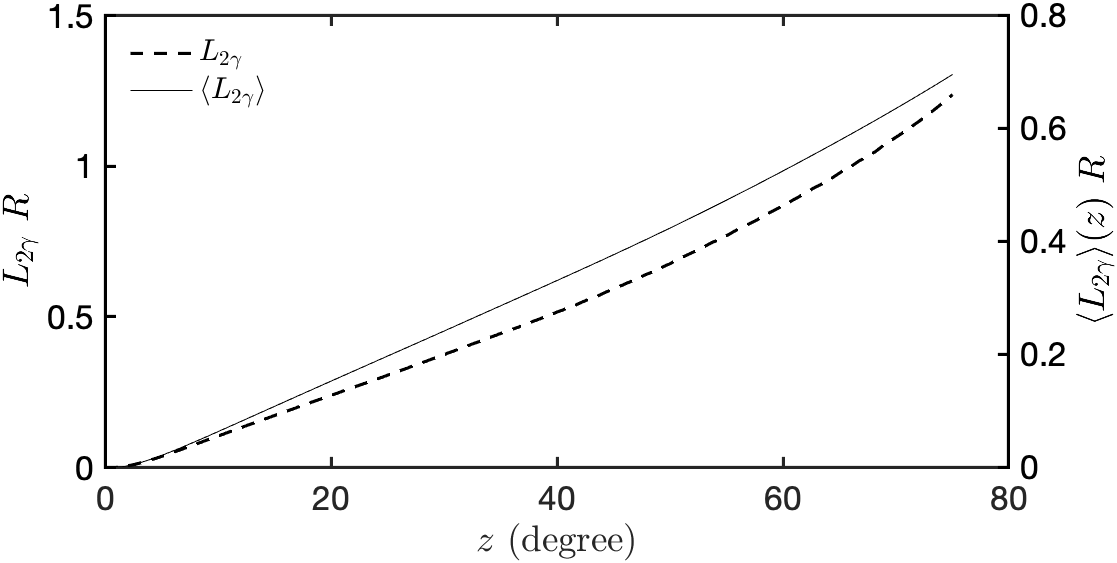}
   \caption{\small ({\it Left}): Geometry of the Exosphere.  The
   coordinate system is defined by $x$ away from the Sun, $y$ towards
   the terrestrial North pole and Earth at the origin. The circles
   are successively: the surface of Earth ($r=R_E$; dark line),
   $r_s=R_E+h$ where $h$ is the height of a satellite (marked ``S");
   radius of  $2R_E$ and $3R_E$.  The impact parameter, $a$, is the
   segment OA.  The umbra cast by Earth is a cone with a depth of
   $215\,R_E$.  We simplify by replacing the umbra-penumbra combination
   with  a cylinder (dashed lines; ``Shadow").  Consider a line-of-sight
   originating from the aforementioned satellite going towards point
   B (which is at radius $r$).   Let $z$ be the zenith angle,
   measured from $x$ axis in a counter-clockwise fashion, and $\phi$
   the corresponding geo-centric polar angle.  $l$ (SB) is the
   distance measured from the satellite to point B.  ({\it Right}): The model
   two-photon surface brightness in units of Rayleigh expected at
   night time as a function of the fraction of the sky, as  measured
   along angle $z$ of the satellite.}
 \label{fig:Exosphere_Geometry}
\end{figure*}

\subsection{Ly$\beta$ scattering during night time}

We now compute Ly$\beta$ scattering during Earth eclipse as observed
by a satellite  in a circular orbit of height $h$, at local midnight
(see Figure~\ref{fig:Exosphere_Geometry} for geometry).  Not much
error is made in replacing the conical umbra with a cylindrical
shadow.  From trigonometry we find $r^2=r_s^2+l^2+2r_s l\cos(z)$
where $r_s=R_E+h\,$ is  orbital radius of the satellite 
and $l$ is the distance measured from the satellite along a given
line-of-sight (see Figure~\ref{fig:Exosphere_Geometry} for geometry).
We integrate the column of hydrogen along a line-of-sight bearing
angle $z$ to obtain $N_H$.  The resulting Ly$\beta$ photon surface
brightness is $L_\beta = (1/4\pi)N_H\mathcal{R}_\beta$ where
$\mathcal{R}_\beta$ is the  per atom rate of Ly$\beta$ excitation
(Equation~\ref{eq:R_beta}). The corresponding two-photon
emission per atom rate is $(1-Bp_{2\gamma})R_\beta$
(see \S\ref{sec:Excitation_2s_Solar}).

In Figure~\ref{fig:Exosphere_Geometry} we display
$L_{2\gamma}$ as a function of the satellite zenith angle. We see
that in our simple single-scattering model there is no two-photon
emission at zenith.  However, there is evidence of Ly$\beta$ photons
in the shadow cylinder. First, the upward looking Ly$\beta$+OI
(\S\ref{sec:LowEarthMissions}) is substantial, $\approx 8\,R$. Next,
O~I $\lambda\,$1304\,\AA\ triplet is seen by HST at local midnight.
This bright line is a result of Bowen fluorescence of O~I powered
by Ly$\beta$ photons (\S\ref{sec:Thermosphere_Appendix}).  Clearly,
Ly$\beta$ must be scattered into the shadow cylinder.

We make the simplistic assumption that the scattered photons  ``fill
up the trough" in Figure~\ref{fig:Exosphere_Geometry}.  \GALEX\
observations were conducted for less than a quarter of the orbit.
Thus the observing window is  $z<45^\circ$.  The mean model brightness
over this window is $0.4\,R$ which corresponds to 34\,CU.

\section{Summing up}
 \label{sec:SummingUp}

\begin{deluxetable}{lllr}[htb]
\label{tab:Inventory}
 \tablecaption{Inventory of  diffuse FUV emission}
 \tablewidth{0pt}
 \tablehead{
 \colhead{Source}&
 \colhead{tracer}&
 \colhead{value}&
 \colhead{$\mathcal{B}$ (CU)}
 }
\startdata
WIM & H$\alpha$ & $0.4\,R$ &  23 \\
HIM & C~IV & 8625\,LU&  34\\
Low-velocity shocks$^{a}$ & model & - &20\\
 IPM$^{b,c}$ & Ly$\beta$ & $2.4\,R$ & 7--28 \\
 Exobase$^d$ & model & ?? & ??\\
Exosphere$^c$  & model & - & 34 \\
\hline
 Total & & &  118--139\\
 \hline
 \hline
 Offset component &\GALEX\ & - &120--180\\
 \enddata
 \tablecomments{\small The source of emission is two-photon for all
 entries save the HIM (for which the source is line emission).
 $^a$This entry is  computed from a theoretical model. Restricted
 to shocks which are at least 100\,km\,s$^{-1}$.  $^b$The intensity
 of the IPM is brightest in the upwind direction and faintest in
 the opposite direction. $^c$This contribution scales directly with
 solar activity (EUV flux). Estimates presented here were computed
 for solar minimum. $^d$This refers to the region with height from
 1200\,km to 1500\,km.}
\end{deluxetable}

We started the paper by noting that at high latitude some 120--180\,CU
of the \GALEX\ FUV could not be accounted by EBL and DGL
(\S\ref{sec:DiffuseFUVEmission}).  In successive sections
(\S\ref{sec:TheGalaxy}-\S\ref{sec:Exosphere}) we undertook a systematic
examination of two-photon emission from the WIM, low-velocity shocks,
and the Solar system as well as line emission from the HIM.  The
resulting estimates are summarized in Table~\ref{tab:Inventory}
above.  We find  that 118--139\,CU of the  \GALEX\ FUV background
can be attributed to conventional sources.  Given the ramifications of
this conclusion we next examine the uncertainties in the entries
in Table~\ref{tab:Inventory}.

\subsection{Uncertainties}

In my view, the contributions from the WIM, the HIM and IPM rest
on reliable measurements and robust theory (WIM: classical theory
of recombination; HIM: direct observations; IPM: basic physics,
optically thin situation and robust measurements of density profile
of IPM H atoms)

By far the biggest uncertainty is the contribution
from the exobase region.  Recall that  H~I dominates over O~I
starting at 700\,km (see \S\ref{sec:Thermosphere_Appendix}) and the
model for the density of H atoms deduced from SWAN or IMAGE is reliable only
above 1500\,km (see \S\ref{sec:SWAN-IMAGE}).  Below height of 1200\,km
the high proton density suppresses two-photon decay
(\S\ref{sec:CollisionalMixing}).  
 {Thus, Ly$\beta$ scattering of H atoms below height
 of 1200\,km results in H$\alpha$ but no two-photon continuum.} 
H~atoms in the annulus between 1200 and 1500\,km will fluoresce and
produce both H$\alpha$ and two-photon continuum. For short-hand we
refer to this region as the exobase.  Since we lack reliable
measurements for atomic hydrogen in this region there is no numerical
entry for exobase in Table~\ref{tab:Inventory}.

 {
 Another notable uncertainty is the contribution of two-photon
 emission from low-velocity shocks. We break this uncertainty in
 two parts.  First, is the uncertainty in the efficiency factor of
 supernova shocks, $\eta$ (see \S\ref{sec:LowVelocityShocks}). }
We  took the average of two estimates, $\eta=[0.02,0.05]$. This
directly translates to  factor of 2.5 uncertainty in the corresponding
entry in Table~\ref{tab:Inventory}.  
 {Next, we remind the reader that  we have not included
 contributions from ``lesser" shocks,
 e.g. stellar winds and novae and spiral density shocks. } 
In this spirit, we acknowledge ignoring contribution to photon
emission from shocks due to infall of gas into the disk.  While the
average infall rate does not compete with supernova shocks the
infall is localized -- the intermediate and high velocity cloud
system \citep{wvw97}.  Along these lines we note that the integrated
spectrum towards the North Celestial Pole (NCP) shows a constant
$N_{\rm HI}(v)v^2$ out to velocity of 80\,km\,s$^{-1}$ and significant
infall \citep{kf85}.  Note that the latter feature is not seen
towards the SCP, another manifestation of localized flows.

Finally, the two-photon emission from the Earth and IPM scales
directly with solar EUV irradiance. This typically varies  by a
factor of 2.5 over a solar cycle.  In computing the contribution
from the exosphere we  adopted the solar minimum value for solar
Ly$\beta$ irradiance (Equation~\ref{eq:F_solar_beta}).

We conclude that a significant fraction, say two-third, and perhaps
even all of the offset component can be accounted for by conventional
sources discussed in this paper.  We cannot rule out
 {that a fraction of the offset emission} is from
 exotic sources such as decay of dark matter but Occam's razor
 suggests otherwise.

\subsection{Global H$\alpha$ constraint}

A question raised by the referee is the constraint on two-photon
emission provided by observations of diffuse H$\alpha$.  Let $X$
be the photon ratio of two-photon emission to H$\alpha$ emission.
For recombination is $X$ is $[0.63, 0.72, 0.85]$ for $T=[5,10,20]\times
10^3\,$K.  For Ly$\beta$ fluorescence in optically thin and low
density medium ($n_p\ll 10^4\,{\rm cm^{-3}}$), $X=1$.  In contrast,
$X\rightarrow 0$ for high density medium (e.g.  thermosphere). HIM
contributes to the FUV emission but has neither an associated
H$\alpha$ emission or two-photon continuum.  
 {In contrast, $X\rightarrow 0$ for post-shocked
 partially ionized ``warm" ($10^4\lesssim T \lesssim 10^5\,$K) gas.
 In such gas, the excitation is primarily to the $n=2$ level and
 so the cooling is via Ly$\alpha$ and two-photon continuum; see
 \citet{ks22} for further cooling curves.}

Subtracting contributions from the HIM and two-photon continuum
exosphere (which will have an associated H$\alpha$ but at zero
topocentric velocity and readily marked as such by WHAM) we find
the minimum two-photon background is 50\,CU (see
Table~\ref{tab:Inventory}).  Using the conversion factor computed
in \S\ref{sec:TwoPhotonDecay} this amounts to a two-photon surface
brightness of 0.58\,$R$. The expected H$\alpha$ surface brightness
is then $\approx 0.58\langle X^{-1}\rangle\,R$ where the angular
brackets stand for averaging over all emission processes.

The only facility that can approach this low surface brightness is from
WHAM.  The observed value at the Galactic poles is $0.5\,R$ (see
\S\ref{sec:WarmIonizedMedium}). However, there are two caveats that
must be noted. First, some of the diffuse H$\alpha$ is reflected
light. This is estimated to be 20\% (see \S\ref{sec:WarmIonizedMedium}).
Next, WHAM relied on velocity separation to subtract geo-coronal
contribution. In effect, the WHAM methodology is well suited to
measuring features which are compact in velocity space. Measuring
H$\alpha$ emission that is wide in velocity would require very
careful accounting of the spectral baseline. Caveats aside, reconciling
the model presented here to the WHAM observations would require
$\langle X^{-1}\approx 0.69\rangle$ which, in our model, means that
collisional excitations contribute more than recombinations. 

\begin{figure}[hbtp]   % exoray_UVEX.m
 \plotone{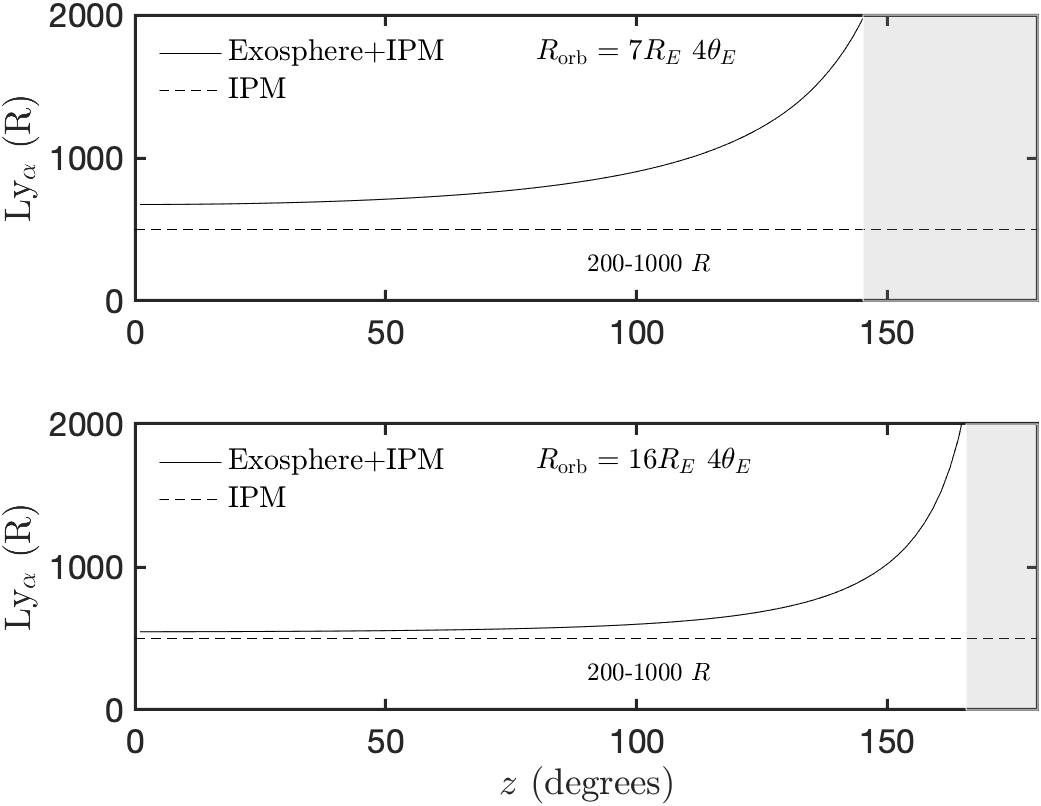}
  \caption{\small The expected Ly$\alpha$ background (in $R$) as a
  function of $z$ for a satellite located in geo-synchronous orbit
  ($6.6\,R_E$; e.g., IUE or the planned Spektr-UF mission;
  \citealt{bss+16}) and in High Earth Orbit (HEO; $16\,R_E$).  The
  shaded region covers the zenith angle range of $180^\circ$ to
  $180^\circ-4\theta_E$ where $\theta_E$ is the angular radius of
  Earth as seen from the vantage point of the satellite.  For the
  purpose of this illustration the IPM Ly$\alpha$ surface brightness
 is fixed to 500\,$R$ (but see \S\ref{sec:IPM}).
  Note that the two-photon continuum tracks the Ly$\alpha$ surface
  brightness.  The two-photon continuum intensity  is equal to ${\rm
  Ly_\alpha}/4000\,R$.  (see \S\ref{sec:Fluorescence_Solar}).}
 \label{fig:HEO}
\end{figure}

\section{Concluding Thoughts}
 \label{sec:ConcludingThoughts}

The diffuse FUV radiation seen at high Galactic latitudes has two
components:  emission that is correlated with cirrus clouds and
emission that is independent of cirrus clouds.  The former is
attributed to stellar FUV light that is reflected by dust particles.
For the latter,  \citet{amr+19} attribute 196-131\,CU to emission
from extra-galactic sources (galaxies, QSOs and IGM). After
accounting for the extragalactic sources some  120--180\,CU of
emission is left -- the so-called  ``offset" component.

This paper investigated FUV emission conventional sources: two-photon
continuum from the Galactic WIM and low-velocity  shocks, FUV line
emission from HIM and two-photon emission our own interplanetary
medium, and exosphere and the thermosphere.  From Table~\ref{tab:Inventory}
it appears that these contributions collectively can readily account
for two thirds of the offset component. We did not compute emission
from a number of minor processes. It is possible that the offset
component can be entirely explained without resorting to speculation
(e.g.\ decay of dark matter).

We conclude this article with a few thoughts. Apart from Ly$\alpha$
the irreducible background for FUV missions is, in our model, due
to two-photon emission. Both backgrounds can be considerably reduced
by placing FUV missions into a high-earth orbit (Figure~\ref{fig:HEO})
and by undertaking observations at satellite zenith angle, $z\lesssim
90^\circ$ {\it and} preferentially studying regions of sky centered
on the down-wind direction.

The discovery of three large SNRs \citep{fdw+21} was made possible,
in part, by the distinctive FUV signature of low-velocity shocks.
Two-photon continuum  is more sensitive than H$\alpha$ searches
because 
(1) the FUV sky is dark and so diffuse emission
is readily detected and (2) H$\alpha$ from low-velocity shocks is
broad in velocity and also has to compete with other sources (e.g.,
geo-corona, IPM and WIM).  
Motivated thus we advocate
for an FUV mission which will undertake an all-sky survey with
arcsecond resolution.  The resulting FUV continuum image of the sky
will provide a new diagnostic for study of diffuse Galactic structures
and low-velocity shocks.  A concomitant effort should be modeling
the collective cooling from the innumerable low-velocity shocks in
our Galaxy.  

For satellites in LEO, the primary airglow in FUV is
due to Ly$\alpha$ and lines of oxygen. The \GALEX\ FUV bandpass was
designed to  avoid the two strongest lines, Ly$\alpha$ and the
fluorescent O~I triplet (see \S\ref{sec:OI_BowenFluorescence}).  A small amount of the
semi-forbidden 1356\,\AA\ oxygen doublet is present within the
bandpass.  The
airglow is the smallest when the satellite is in the shadow of Earth
and is pointing anti-sun.   This airglow is seen by FUV instruments\footnote{see
Figure 7.1 of the COS instrument handbook. \url{https://hst-docs.stsci.edu/cosihb}} on HST
even when pointing anti-sun, whilst in deep Earth shadow. An  FUV mission 
in HEO would not suffer from airglow and so the background emission
will be very stable. A happy by-product of such a mission would be an
accurate measurement of the strength of the offset component.

Next, from the discussion in \S\ref{sec:Thermosphere} it is clear
that there is a tension between the amount of hydrogen in the
thermosphere as deduced by modeling ground-based H$\alpha$ data and
the MSIS model. Furthermore, the modeling of LEO-based measurements
of Ly$\beta$+OI, complicated by Bowen fluorescence of O~I, is not
satisfactory.  A cubesat with an H$\alpha$ photometer offers a clean
way to model the vertical distribution of hydrogen.

\begin{figure}[tp]
 \centering
  \plotone{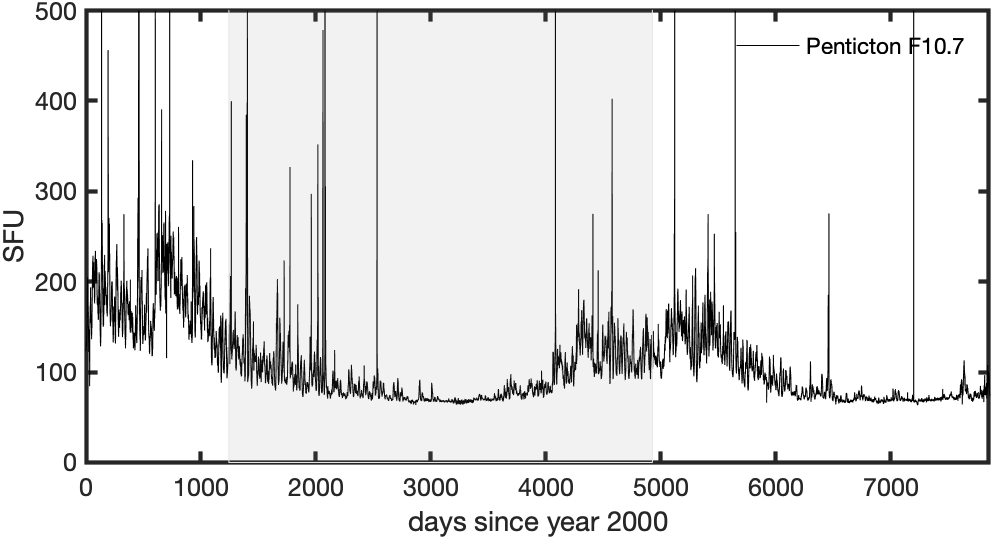}  %Penticton_SFU.m
   \caption{\small The solar 10.7-cm flux density (F10.7) since
   year 2000.  The F10.7 index is widely used as a ground-based
   surrogate for the EUV irradiance of the Sun (see \citealt{T13}).
   The solar minimum occurred in 2010 (say, day 3600).  The shaded
   region is the period during which \GALEX\ was operational.  The
   vertical axis has been clipped.  The short spikes can reach
   2000\,SUF and are probably due to magnetic reconnection.  The
   unit SFU stands for solar flux unit and is equal to $10^4\,$Jy.
   }
 \label{fig:Penticton_Solar}
\end{figure}

As can be seen from Figure~\ref{fig:Penticton_Solar}\footnote{The
measurements were obtained at the Dominion Radio Astrophysical
Observatory, Penticton, British Columbia, Canada and the data were
obtained from \url{https://lasp.colorado.edu/lisird/}.} the solar
EUV irradiance varies on timescales of not just solar cycles but
hours!  This variability offers us an opportunity to  probe
the exosphere and the IPM.  Parenthetically, we note that evidence
for variability in the \GALEX\ FUV data can be readily seen in
discordant ``sky" values for adjacent pointings of GALEX FUV mosaic
images  (e.g., see Figure~2 of \citealt{fdw+21}). Future FUV missions
would be well advised to model the FUV background by folding in
solar EUV monitoring data.

Finally the ISM-Solar system interaction is a major area of research
in space sciences. For a modest funding,  a ground-based facility
with sensitivity ten times better than that of WHAM can be built.
Solar EUV irradiance changes on a variety of timescales.  With this
facility,  it may be possible to measure the resulting H$\alpha$
variations  and thereby ``plumb" the three-dimensional distribution
of the IPM.

\acknowledgments

I gratefully acknowledge receiving help from Ronald Reynolds; Michael
Shull, University of Colorado at Boulder; E. Sterl Phinney, California
Institute of Technology (Caltech); Jayant Murthy, Indian Institute
of Astrophysics;  Edwin J.  Mierkiewicz \&\ Matthew D.\ Zettergren,
Embry-Riddle Aeronautical University; I.\ I.\ Baliukin, Space
Research Institute (IKI), Moscow; and E.\ C.\ Stone, Project Scientist
for Voyager Mission; Nikolaus Zen Prusinski, Caltech.  For feedback:
Ilaria Caiazzo, Caltech; Bruce Draine, Princeton University; Jerry
Edelstein \&\ Christopher McKee, University of California at Berkeley;
Marc Postman, Space Telescope Science Institute; and
Kevian Stassun, Vanderbilt University, Finally, I am most grateful
to Robert Benjamin,University of Wisconsin at Madison; Hannah
Earnshaw, Caltech; Eran Ofek, Weizmann, Institute of Science; and
Michael Shull for careful reading and constructive feedback.
{
Finally, I thank the referee (anonymous) for feedback which resulted
in a clearer and shorter paper.}

\bibliography{bibFUVB}{}

\begin{thebibliography}{}
\expandafter\ifx\csname natexlab\endcsname\relax\def\natexlab#1{#1}\fi

\bibitem[{{Ake}(2012)}]{A12}
{Ake}, T.~B. 2012, in American Astronomical Society Meeting Abstracts, Vol.
  219, American Astronomical Society Meeting Abstracts \#219, 241.18

\bibitem[{{Akshaya} {et~al.}(2018){Akshaya}, {Murthy}, {Ravichandran}, {Henry},
  \& {Overduin}}]{amr+18}
{Akshaya}, M.~S., {Murthy}, J., {Ravichandran}, S., {Henry}, R.~C., \&
  {Overduin}, J. 2018, \apj, 858, 101

\bibitem[{{Akshaya} {et~al.}(2019){Akshaya}, {Murthy}, {Ravichandran}, {Henry},
  \& {Overduin}}]{amr+19}
---. 2019, \mnras, 489, 1120

\bibitem[{{Baliukin} {et~al.}(2019){Baliukin}, {Bertaux}, {Qu{\'e}merais},
  {Izmodenov}, \& {Schmidt}}]{bbq+19}
{Baliukin}, I.~I., {Bertaux}, J.~L., {Qu{\'e}merais}, E., {Izmodenov}, V.~V.,
  \& {Schmidt}, W. 2019, Journal of Geophysical Research (Space Physics), 124,
  861

\bibitem[{{Barnes} {et~al.}(2011){Barnes}, {van Zee}, \& {Skillman}}]{bls11}
{Barnes}, K.~L., {van Zee}, L., \& {Skillman}, E.~D. 2011, \apj, 743, 137

\bibitem[{{Bertaux} \& {Blamont}(1971)}]{bb71}
{Bertaux}, J.~L., \& {Blamont}, J.~E. 1971, \aap, 11, 200

\bibitem[{{Bertaux} {et~al.}(1995){Bertaux}, {Kyr{\"o}l{\"a}}, {Qu{\'e}merais},
  {Pellinen}, {Lallement}, {Schmidt}, {Berth{\'e}}, {Dimarellis}, {Goutail},
  {Taulemesse}, {Bernard}, {Leppelmeier}, {Summanen}, {Hannula}, {Huomo},
  {Kehl{\"a}}, {Korpela}, {Lepp{\"a}l{\"a}}, {Str{\"o}mmer}, {Torsti},
  {Viherkanto}, {Hochedez}, {Chretiennot}, {Peyroux}, \& {Holzer}}]{bkq+95}
{Bertaux}, J.~L., {Kyr{\"o}l{\"a}}, E., {Qu{\'e}merais}, E., {et~al.} 1995,
  \solphys, 162, 403

\bibitem[{{Bishop} {et~al.}(2001){Bishop}, {Harlander}, {Nossal}, \&
  {Roesler}}]{bhn+01}
{Bishop}, J., {Harlander}, J., {Nossal}, S., \& {Roesler}, F.~L. 2001, Journal
  of Atmospheric and Solar-Terrestrial Physics, 63, 341

\bibitem[{{Bowen} {et~al.}(2008){Bowen}, {Jenkins}, {Tripp}, {Sembach},
  {Savage}, {Moos}, {Oegerle}, {Friedman}, {Gry}, {Kruk}, {Murphy}, {Sankrit},
  {Shull}, {Sonneborn}, \& {York}}]{bjt+08}
{Bowen}, D.~V., {Jenkins}, E.~B., {Tripp}, T.~M., {et~al.} 2008, \apjs, 176, 59

\bibitem[{{Boyarchuk} {et~al.}(2016){Boyarchuk}, {Shustov}, {Savanov},
  {Sachkov}, {Bisikalo}, {Mashonkina}, {Wiebe}, {Shematovich}, {Shchekinov},
  {Ryabchikova}, {Chugai}, {Ivanov}, {Voshchinnikov}, {Gomez de Castro},
  {Lamzin}, {Piskunov}, {Ayres}, {Strassmeier}, {Jeffrey}, {Zwintz}, {Shulyak},
  {G{\'e}rard}, {Hubert}, {Fossati}, {Lammer}, {Werner}, {Zhilkin},
  {Kaigorodov}, {Sichevskii}, {Ustamuich}, {Kanev}, \& {Kil'pio}}]{bss+16}
{Boyarchuk}, A.~A., {Shustov}, B.~M., {Savanov}, I.~S., {et~al.} 2016,
  Astronomy Reports, 60, 1

\bibitem[{{Bracco} {et~al.}(2020){Bracco}, {Benjamin}, {Alves}, {Lehmann},
  {Boulanger}, {Montier}, {Mittelman}, {di Cicco}, \& {Walker}}]{bba+20}
{Bracco}, A., {Benjamin}, R.~A., {Alves}, M.~I.~R., {et~al.} 2020, \aap, 636,
  L8

\bibitem[{{Brugel} {et~al.}(1982){Brugel}, {Shull}, \& {Seab}}]{bss82}
{Brugel}, E.~W., {Shull}, J.~M., \& {Seab}, C.~G. 1982, \apjl, 262, L35

\bibitem[{{Chakrabarti} {et~al.}(1984){Chakrabarti}, {Kimble}, \&
  {Bowyer}}]{ckb84}
{Chakrabarti}, S., {Kimble}, R., \& {Bowyer}, S. 1984, \jgr, 89, 5660

\bibitem[{{Chance} \& {Kurucz}(2010)}]{ck10}
{Chance}, K., \& {Kurucz}, R.~L. 2010, \jqsrt, 111, 1289

\bibitem[{{Deharveng} {et~al.}(1982){Deharveng}, {Joubert}, \& {Barge}}]{djb82}
{Deharveng}, J.~M., {Joubert}, M., \& {Barge}, P. 1982, \aap, 109, 179

\bibitem[{{Del Zanna} {et~al.}(2021){Del Zanna}, {Dere}, {Young}, \&
  {Landi}}]{ddy+21}
{Del Zanna}, G., {Dere}, K.~P., {Young}, P.~R., \& {Landi}, E. 2021, \apj, 909,
  38

\bibitem[{{Dennison} {et~al.}(2005){Dennison}, {Turner}, \& {Minter}}]{dtm05}
{Dennison}, B., {Turner}, B.~E., \& {Minter}, A.~H. 2005, \apj, 633, 309

\bibitem[{{Dere} {et~al.}(1997){Dere}, {Landi}, {Mason}, {Monsignori Fossi}, \&
  {Young}}]{dlm+97}
{Dere}, K.~P., {Landi}, E., {Mason}, H.~E., {Monsignori Fossi}, B.~C., \&
  {Young}, P.~R. 1997, \aaps, 125, 149

\bibitem[{{Dong} \& {Draine}(2011)}]{dd11}
{Dong}, R., \& {Draine}, B.~T. 2011, \apj, 727, 35

\bibitem[{{Dopita} {et~al.}(1982){Dopita}, {Binette}, \& {Schwartz}}]{dbs82}
{Dopita}, M.~A., {Binette}, L., \& {Schwartz}, R.~D. 1982, \apj, 261, 183

\bibitem[{{Dove} \& {Shull}(1994)}]{ds94}
{Dove}, J.~B., \& {Shull}, J.~M. 1994, \apj, 430, 222

\bibitem[{{Draine}(2011)}]{D11}
{Draine}, B.~T. 2011, {Physics of the Interstellar and Intergalactic Medium}
  (Princeton University Press)

\bibitem[{{Draine} \& {Bertoldi}(1996)}]{db96}
{Draine}, B.~T., \& {Bertoldi}, F. 1996, \apj, 468, 269

\bibitem[{{Drake} {et~al.}(1969){Drake}, {Victor}, \& {Dalgarno}}]{dvd69}
{Drake}, G.~W., {Victor}, G.~A., \& {Dalgarno}, A. 1969, Physical Review, 180,
  25

\bibitem[{{Dudok de Wit} {et~al.}(2005){Dudok de Wit}, {Lilensten},
  {Aboudarham}, {Amblard}, \& {Kretzschmar}}]{dla+05}
{Dudok de Wit}, T., {Lilensten}, J., {Aboudarham}, J., {Amblard}, P.~O., \&
  {Kretzschmar}, M. 2005, Annales Geophysicae, 23, 3055

\bibitem[{{Duley} \& {Williams}(1980)}]{dw80}
{Duley}, W.~W., \& {Williams}, D.~A. 1980, \apjl, 242, L179

\bibitem[{{Edelstein} {et~al.}(2006){Edelstein}, {Korpela}, {Adolfo}, {Bowen},
  {Feuerstein}, {Hull}, {Jelinsky}, {Nishikida}, {McKee}, {Berg}, {Chung},
  {Fischer}, {Min}, {Oh}, {Rhee}, {Ryu}, {Shinn}, {Han}, {Jin}, {Lee}, {Nam},
  {Park}, {Seon}, \& {Yuk}}]{eka+06}
{Edelstein}, J., {Korpela}, E.~J., {Adolfo}, J., {et~al.} 2006, \apjl, 644,
  L159

\bibitem[{{Fesen} {et~al.}(2021){Fesen}, {Drechsler}, {Weil}, {Strottner},
  {Raymond}, {Rupert}, {Milisavljevic}, {Subrayan}, {di Cicco}, {Walker},
  {Mittelman}, \& {Ludgate}}]{fdw+21}
{Fesen}, R.~A., {Drechsler}, M., {Weil}, K.~E., {et~al.} 2021, arXiv e-prints,
  arXiv:2102.12599

\bibitem[{{Frisch} {et~al.}(2011){Frisch}, {Redfield}, \& {Slavin}}]{frs11}
{Frisch}, P.~C., {Redfield}, S., \& {Slavin}, J.~D. 2011, \araa, 49, 237

\bibitem[{{Frisch} {et~al.}(2013){Frisch}, {Bzowski}, {Livadiotis}, {McComas},
  {Moebius}, {Mueller}, {Pryor}, {Schwadron}, {Sok{\'o}{\l}}, {Vallerga}, \&
  {Ajello}}]{pbl+13}
{Frisch}, P.~C., {Bzowski}, M., {Livadiotis}, G., {et~al.} 2013, Science, 341,
  1080

\bibitem[{{Gladstone} {et~al.}(2018){Gladstone}, {Pryor}, {Stern}, {Ennico},
  {Olkin}, {Spencer}, {Weaver}, {Young}, {Bagenal}, {Cheng}, {Cunningham},
  {Elliott}, {Greathouse}, {Hinson}, {Kammer}, {Linscott}, {Parker},
  {Retherford}, {Steffl}, {Strobel}, {Summers}, {Throop}, {Versteeg}, \&
  {Davis}}]{gps+18}
{Gladstone}, G.~R., {Pryor}, W.~R., {Stern}, S.~A., {et~al.} 2018, \grl, 45,
  8022

\bibitem[{{Gladstone} {et~al.}(2021){Gladstone}, {Pryor}, {Hall}, {Kammer},
  {Strobel}, {Weaver}, {Spencer}, {Retherford}, {Versteeg}, {Davis}, {Young},
  {Steffl}, {Parker}, {Lisse}, {Singer}, \& {Stern}}]{gph+21}
{Gladstone}, G.~R., {Pryor}, W.~R., {Hall}, D.~T., {et~al.} 2021, \aj, 162, 241

\bibitem[{{Gry} \& {Jenkins}(2017)}]{gj17}
{Gry}, C., \& {Jenkins}, E.~B. 2017, \aap, 598, A31

\bibitem[{{Haffner} {et~al.}(2003){Haffner}, {Reynolds}, {Tufte}, {Madsen},
  {Jaehnig}, \& {Percival}}]{hrt+03}
{Haffner}, L.~M., {Reynolds}, R.~J., {Tufte}, S.~L., {et~al.} 2003, \apjs, 149,
  405

\bibitem[{{Haffner} {et~al.}(2009){Haffner}, {Dettmar}, {Beckman}, {Wood},
  {Slavin}, {Giammanco}, {Madsen}, {Zurita}, \& {Reynolds}}]{hdb+09}
{Haffner}, L.~M., {Dettmar}, R.~J., {Beckman}, J.~E., {et~al.} 2009, Reviews of
  Modern Physics, 81, 969

\bibitem[{{Haffner} {et~al.}(2010){Haffner}, {Reynolds}, {Madsen}, {Hill},
  {Barger}, {Jaehnig}, {Mierkiewicz}, {Percival}, \& {Chopra}}]{hrm+10}
{Haffner}, L.~M., {Reynolds}, R.~J., {Madsen}, G.~J., {et~al.} 2010, in
  Astronomical Society of the Pacific Conference Series, Vol. 438, The Dynamic
  Interstellar Medium: A Celebration of the Canadian Galactic Plane Survey, ed.
  R.~{Kothes}, T.~L. {Landecker}, \& A.~G. {Willis}, 388

\bibitem[{{Hamden} {et~al.}(2013){Hamden}, {Schiminovich}, \&
  {Seibert}}]{hss13}
{Hamden}, E.~T., {Schiminovich}, D., \& {Seibert}, M. 2013, \apj, 779, 180

\bibitem[{{Henry}(1991)}]{H91}
{Henry}, R.~C. 1991, \araa, 29, 89

\bibitem[{{Henry} {et~al.}(2015){Henry}, {Murthy}, {Overduin}, \&
  {Tyler}}]{hmo+15}
{Henry}, R.~C., {Murthy}, J., {Overduin}, J., \& {Tyler}, J. 2015, \apj, 798,
  14

\bibitem[{{Holberg}(1986)}]{H86}
{Holberg}, J.~B. 1986, \apj, 311, 969

\bibitem[{{Jo} {et~al.}(2017){Jo}, {Seon}, {Min}, {Edelstein}, \&
  {Han}}]{jsm+17}
{Jo}, Y.-S., {Seon}, K.-I., {Min}, K.-W., {Edelstein}, J., \& {Han}, W. 2017,
  \apjs, 231, 21

\bibitem[{{Jo} {et~al.}(2019){Jo}, {Seon}, {Min}, {Edelstein}, {Han},
  {Korpela}, \& {Sirk}}]{jsm+19}
{Jo}, Y.-S., {Seon}, K.-i., {Min}, K.-W., {et~al.} 2019, \apjs, 243, 9

\bibitem[{{Jura}(1974)}]{J74}
{Jura}, M. 1974, \apj, 191, 375

\bibitem[{{Kim} \& {Ostriker}(2015)}]{ko15}
{Kim}, C.-G., \& {Ostriker}, E.~C. 2015, \apj, 802, 99

\bibitem[{{Kollmeier} {et~al.}(2014){Kollmeier}, {Weinberg}, {Oppenheimer},
  {Haardt}, {Katz}, {Dav{\'e}}, {Fardal}, {Madau}, {Danforth}, {Ford},
  {Peeples}, \& {McEwen}}]{kwo+14}
{Kollmeier}, J.~A., {Weinberg}, D.~H., {Oppenheimer}, B.~D., {et~al.} 2014,
  \apjl, 789, L32

\bibitem[{{Kulkarni} \& {Fich}(1985)}]{kf85}
{Kulkarni}, S.~R., \& {Fich}, M. 1985, \apj, 289, 792

\bibitem[{{Kulkarni} \& {Shull}(2022)}]{ks22}
{Kulkarni}, S.~R., \& {Shull}, M.~J. 2022, manuscript in prep

\bibitem[{{Kurt} \& {Sunyaev}(1970)}]{ks70}
{Kurt}, V.~G., \& {Sunyaev}, R.~A. 1970, in Ultraviolet Stellar Spectra and
  Related Ground-Based Observations, ed. R.~{Muller}, L.~{Houziaux}, \& H.~E.
  {Butler}, Vol.~36, 341

\bibitem[{{Lemaire} {et~al.}(2015){Lemaire}, {Vial}, {Curdt}, {Sch{\"u}hle}, \&
  {Wilhelm}}]{lvc+15}
{Lemaire}, P., {Vial}, J.~C., {Curdt}, W., {Sch{\"u}hle}, U., \& {Wilhelm}, K.
  2015, \aap, 581, A26

\bibitem[{{Lemaire} {et~al.}(2012){Lemaire}, {Vial}, {Curdt}, {Sch{\"u}hle}, \&
  {Woods}}]{lvc+12}
{Lemaire}, P., {Vial}, J.~C., {Curdt}, W., {Sch{\"u}hle}, U., \& {Woods}, T.~N.
  2012, \aap, 542, L25

\bibitem[{{Lockman} \& {Gehman}(1991)}]{lg91}
{Lockman}, F.~J., \& {Gehman}, C.~S. 1991, \apj, 382, 182

\bibitem[{{L{\'o}pez-Moreno} {et~al.}(2001){L{\'o}pez-Moreno}, {Morales},
  {G{\'o}mez}, {Trapero}, {Bowyer}, {Edelstein}, {Korpela}, \&
  {Lampton}}]{lmg+01}
{L{\'o}pez-Moreno}, J.~J., {Morales}, C., {G{\'o}mez}, J.~F., {et~al.} 2001,
  \apss, 276, 211

\bibitem[{{Manchester} \& {Taylor}(1977)}]{mt77}
{Manchester}, R.~N., \& {Taylor}, J.~H. 1977, {Pulsars} (W. H. Freeman \& Co)

\bibitem[{{Martin} {et~al.}(1990){Martin}, {Hurwitz}, \& {Bowyer}}]{mhb90}
{Martin}, C., {Hurwitz}, M., \& {Bowyer}, S. 1990, \apj, 354, 220

\bibitem[{{Martin} {et~al.}(1991){Martin}, {Hurwitz}, \& {Bowyer}}]{mhb91}
---. 1991, \apj, 379, 549

\bibitem[{{Martin} {et~al.}(2005){Martin}, {Fanson}, {Schiminovich},
  {Morrissey}, {Friedman}, {Barlow}, {Conrow}, {Grange}, {Jelinsky},
  {Milliard}, {Siegmund}, {Bianchi}, {Byun}, {Donas}, {Forster}, {Heckman},
  {Lee}, {Madore}, {Malina}, {Neff}, {Rich}, {Small}, {Surber}, {Szalay},
  {Welsh}, \& {Wyder}}]{mfs+05}
{Martin}, D.~C., {Fanson}, J., {Schiminovich}, D., {et~al.} 2005, \apjl, 619,
  L1

\bibitem[{{McCullough} \& {Benjamin}(2001)}]{mb01}
{McCullough}, P.~R., \& {Benjamin}, R.~A. 2001, \aj, 122, 1500

\bibitem[{{McKee} \& {Ostriker}(1977)}]{mo77}
{McKee}, C.~F., \& {Ostriker}, J.~P. 1977, \apj, 218, 148

\bibitem[{{Meftah} {et~al.}(2021){Meftah}, {Snow}, {Dam{\'e}}, {Bolse{\'e}},
  {Pereira}, {Cessateur}, {Bekki}, {Keckhut}, {Sarkissian}, \&
  {Hauchecorne}}]{msd+21}
{Meftah}, M., {Snow}, M., {Dam{\'e}}, L., {et~al.} 2021, \aap, 645, A2

\bibitem[{{Meier} {et~al.}(1987){Meier}, {Anderson}, {Paxton}, {McCoy}, \&
  {Chakrabarti}}]{map+87}
{Meier}, R.~R., {Anderson}, D.~E., J., {Paxton}, L.~J., {McCoy}, R.~P., \&
  {Chakrabarti}, S. 1987, \jgr, 92, 8767

\bibitem[{{Mende} {et~al.}(2000){Mende}, {Heetderks}, {Frey}, {Stock},
  {Lampton}, {Geller}, {Abiad}, {Siegmund}, {Habraken}, {Renotte}, {Jamar},
  {Rochus}, {Gerard}, {Sigler}, \& {Lauche}}]{mhf+00}
{Mende}, S.~B., {Heetderks}, H., {Frey}, H.~U., {et~al.} 2000, \ssr, 91, 287

\bibitem[{{Miller} \& {Cox}(1993)}]{mc93}
{Miller}, Walter~Warren, I., \& {Cox}, D.~P. 1993, \apj, 417, 579

\bibitem[{{Morales} {et~al.}(1998){Morales}, {G{\'o}mez}, {Trapero}, {Bowyer},
  {Edelstein}, \& {Korpela}}]{mgt+98}
{Morales}, C., {G{\'o}mez}, J.~F., {Trapero}, J., {et~al.} 1998, \apss, 263,
  393

\bibitem[{{Morrissey} {et~al.}(2007){Morrissey}, {Conrow}, {Barlow}, {Small},
  {Seibert}, {Wyder}, {Budav{\'a}ri}, {Arnouts}, {Friedman}, {Forster},
  {Martin}, {Neff}, {Schiminovich}, {Bianchi}, {Donas}, {Heckman}, {Lee},
  {Madore}, {Milliard}, {Rich}, {Szalay}, {Welsh}, \& {Yi}}]{mcb+07}
{Morrissey}, P., {Conrow}, T., {Barlow}, T.~A., {et~al.} 2007, \apjs, 173, 682

\bibitem[{{Murthy} {et~al.}(2010){Murthy}, {Henry}, \& {Sujatha}}]{mhs10}
{Murthy}, J., {Henry}, R.~C., \& {Sujatha}, N.~V. 2010, \apj, 724, 1389

\bibitem[{{Nahar}(2021)}]{N21}
{Nahar}, S.~N. 2021, Atoms, 9, 73

\bibitem[{{Nossal} {et~al.}(2008){Nossal}, {Mierkiewicz}, {Roesler}, {Haffner},
  {Reynolds}, \& {Woodward}}]{nmr+08}
{Nossal}, S.~M., {Mierkiewicz}, E.~J., {Roesler}, F.~L., {et~al.} 2008, Journal
  of Geophysical Research (Space Physics), 113, A11307

\bibitem[{{Nossal} {et~al.}(2019){Nossal}, {Mierkiewicz}, {Roesler},
  {Woodward}, {Gardner}, \& {Haffner}}]{nmr+19}
---. 2019, Journal of Geophysical Research (Space Physics), 124, 10,674

\bibitem[{{Nussbaumer} \& {Schmutz}(1984)}]{ns84}
{Nussbaumer}, H., \& {Schmutz}, W. 1984, \aap, 138, 495

\bibitem[{{O'Connell}(1987)}]{O87}
{O'Connell}, R.~W. 1987, \aj, 94, 876

\bibitem[{{Osterbrock}(1974)}]{O74}
{Osterbrock}, D.~E. 1974, {Astrophysics of gaseous nebulae} (W.\ H.\ Freeman
  and Company)

\bibitem[{{{\O}Stgaard} {et~al.}(2003){{\O}Stgaard}, {Mende}, {Frey},
  {Gladstone}, \& {Lauche}}]{omf+03}
{{\O}Stgaard}, N., {Mende}, S.~B., {Frey}, H.~U., {Gladstone}, G.~R., \&
  {Lauche}, H. 2003, Journal of Geophysical Research (Space Physics), 108, 1300

\bibitem[{{Paresce} \& {Jakobsen}(1980)}]{pj80}
{Paresce}, F., \& {Jakobsen}, P. 1980, \nat, 288, 119

\bibitem[{{Paresce} {et~al.}(1980){Paresce}, {McKee}, \& {Bowyer}}]{pmb80}
{Paresce}, F., {McKee}, C.~F., \& {Bowyer}, S. 1980, \apj, 240, 387

\bibitem[{{Pr{\"o}lss} \& {Bird}(2004)}]{pb04}
{Pr{\"o}lss}, G.~W., \& {Bird}, M.~K. 2004, {Physics of the Earth's Space
  Environment: an introduction} (Springer)

\bibitem[{{Purcell}(1952)}]{P52}
{Purcell}, E.~M. 1952, \apj, 116, 457

\bibitem[{{Raymond}(1979)}]{R79}
{Raymond}, J.~C. 1979, \apjs, 39, 1

\bibitem[{{Reynolds}(1990)}]{R90b}
{Reynolds}, R.~J. 1990, \apjl, 349, L17

\bibitem[{{Reynolds}(1992)}]{R92}
---. 1992, \apjl, 392, L35

\bibitem[{{Sandel} {et~al.}(1978){Sandel}, {Shemansky}, \& {Broadfoot}}]{ssb78}
{Sandel}, B.~R., {Shemansky}, D.~E., \& {Broadfoot}, A.~L. 1978, \nat, 274, 666

\bibitem[{{Savage} {et~al.}(2000){Savage}, {Sembach}, {Jenkins}, {Shull},
  {York}, {Sonneborn}, {Moos}, {Friedman}, {Green}, {Oegerle}, {Blair}, {Kruk},
  \& {Murphy}}]{ssj+00}
{Savage}, B.~D., {Sembach}, K.~R., {Jenkins}, E.~B., {et~al.} 2000, \apjl, 538,
  L27

\bibitem[{{Sciama}(1990)}]{S90}
{Sciama}, D.~W. 1990, \apj, 364, 549

\bibitem[{{Seon} {et~al.}(2011){Seon}, {Edelstein}, {Korpela}, {Witt}, {Min},
  {Han}, {Shinn}, {Kim}, \& {Park}}]{sek+11}
{Seon}, K.-I., {Edelstein}, J., {Korpela}, E., {et~al.} 2011, \apjs, 196, 15

\bibitem[{{Shih} {et~al.}(1984){Shih}, {Scherb}, \& {Roesler}}]{ssr84}
{Shih}, P., {Scherb}, F., \& {Roesler}, F.~L. 1984, \apj, 279, 453

\bibitem[{{Shull} \& {McKee}(1979)}]{sm79}
{Shull}, J.~M., \& {McKee}, C.~F. 1979, \apj, 227, 131

\bibitem[{{Slavin} {et~al.}(2000){Slavin}, {McKee}, \& {Hollenbach}}]{smh00}
{Slavin}, J.~D., {McKee}, C.~F., \& {Hollenbach}, D.~J. 2000, \apj, 541, 218

\bibitem[{{Spitzer} \& {Greenstein}(1951)}]{sg51}
{Spitzer}, Lyman, J., \& {Greenstein}, J.~L. 1951, \apj, 114, 407

\bibitem[{{Spitzer}(1978)}]{S78}
{Spitzer}, L. 1978, {Physical processes in the interstellar medium}
  (Wiley-Interscience), doi:10.1002/9783527617722

\bibitem[{{Stecker}(1980)}]{S80}
{Stecker}, F.~W. 1980, \prl, 45, 1460

\bibitem[{{Sutherland} \& {Dopita}(1993)}]{sd93}
{Sutherland}, R.~S., \& {Dopita}, M.~A. 1993, \apjs, 88, 253

\bibitem[{{Tapping}(2013)}]{T13}
{Tapping}, K.~F. 2013, Space Weather, 11, 394

\bibitem[{{Thomas} \& {Krassa}(1971)}]{tk71}
{Thomas}, G.~E., \& {Krassa}, R.~F. 1971, \aap, 11, 218

\bibitem[{{Wakker} \& {van Woerden}(1997)}]{wvw97}
{Wakker}, B.~P., \& {van Woerden}, H. 1997, \araa, 35, 217

\bibitem[{{Witt} {et~al.}(2010){Witt}, {Gold}, {Barnes}, {DeRoo}, {Vijh}, \&
  {Madsen}}]{wgb+10}
{Witt}, A.~N., {Gold}, B., {Barnes}, Frank~S., I., {et~al.} 2010, \apj, 724,
  1551

\bibitem[{{Woolley}(1934)}]{W34}
{Woolley}, R.~V.~D.~R. 1934, \mnras, 95, 101

\bibitem[{{Zettergren} \& {Snively}(2015)}]{zs15}
{Zettergren}, M.~D., \& {Snively}, J.~B. 2015, Journal of Geophysical Research
  (Space Physics), 120, 8002

\end{thebibliography}
\bibliographystyle{aasjournal}

\begin{comment}
{
 A separate concern is that the 2s to 2p collisional coefficients
 that we used were computed with asymptotic formalism that applies
 for particles moving at velocities  $\gtrsim 10\,{\rm km\,s^{-1}}$
 \citep{P52}. The thermal speeds of protons in the exosphere is
 smaller, $\approx 4\, {\rm km\,s^{-1}}$. We used an empirical model
 (power law fit to the cross-section as a function of temperature)
 to extrapolate  the collisional coefficients at the temperature
 of the exosphere. A definitive calculation of the coefficients at
 $T\approx 10^3\,$K would be reassuring.}
\end{comment}

\appendix
\section{Two-Photon Continuum}
 \label{sec:Two-PhotonContinuum}

The classical reference for the two-photon spectrum is \citet{sg51}
where they invoked the two-photon process to explain the continuum
of planetary nebulae.  The modern reference is
\citet{dvd69}.  For the A-coefficient
we use the fitting formula of \citet{ns84}:
 \begin{eqnarray}
  A(y) &=& CY[1-(4Y)^\gamma] + \alpha C Y^\beta (4Y)^\gamma
   \label{eq:Ay}
 \end{eqnarray}
where  $y=\nu/\nu_0$, 
$Y=y(1-y)$, $\nu_0=c/\lambda_0$
and $\lambda_0=1215.67$\,\AA\ is the wavelength of Ly$\alpha$. The
fit values are $\alpha=0.88$, $\beta=1.53$, $\gamma=0.8$ and
$C=202.0\,{\rm s^{-1}}$.  Since each de-excitation results in the
emission of two photons, $\int_0^1 A(y)dy=2A_{21}=16.4\,{\rm s^{-1}}$.
Noting that $A(y)$ is the probability of emitting a photon in the
frequency interval $dy=d\nu/\nu_0$ we find the emissivity from  a
single decay to be
 \begin{equation}
  j_\nu = \frac{1}{4\pi} \frac{h\nu}{\nu_0}A(y).
	\label{eq:j_nu}
 \end{equation}
Traditionally, observers use the spectral intensity as a function
of wavelength, $j_\lambda=j_\nu\nu/\lambda$.  The corresponding
photon intensity is $n(\lambda)=j_\lambda/(hc/\lambda) \propto \nu
j_\nu$.  While $A(y)$ peaks at $y=1/2$ corresponding to $\nu=\nu_0/2$
the spectral intensity  $j_\lambda$ peaks at 1400\,\AA\
(Figure~\ref{fig:TwoPhoton_Galex}).

\begin{deluxetable}{lrrr}[htb]
 \tablecaption{\GALEX\ bands \& response to two-photon decay}
  \tablewidth{0pt}
   \tablehead{
    \colhead{Parameter}& 
     \colhead{unit} &
     \colhead{FUV}&
     \colhead{NUV}}
\startdata 
Bandpass  &\AA\ & 1350--1750 & 1750--2800 \\
FoV & deg$^2$ 	& 1.267 & 1.227 \\
$\mathcal{I}$  &cm$^2$\AA   & 9402  & 45008 \\
$\Delta\lambda_{\rm eq}$ &\AA\ & 255  & 730 \\
\hline 
$n(2\gamma)$ & count     & 0.27 &  0.44 \\
CR & $ {\rm phot\,s^{-1}}$ & 309 & 838 \\
$\mathcal{B}_{2\gamma}$        & CU &85.1& 18.4 \\
\enddata
 \tablecomments{The vital statistics of the FUV and NUV channels
 are summarized in the top half of the table.  Here, $\mathcal{I}=\int
 A_{\rm eff}(\lambda)d\lambda$ and $\Delta\lambda_{\rm eq}\equiv
 \mathcal{I}/{\rm max}(A_{\rm eff})$.  $n(2\gamma)$ is the number
 of  photons detected in each channel for a single two-photon decay.
 The last two lines are the response by the two \GALEX\ channels
 to a uniform background from a column of $10^6\,{\rm
 decays\,cm^2\,s^{-1}}$.  CR is the counting rate across the entire
 detector for such a background while $\mathcal{B}_{2\gamma}$ is the
 inferred surface brightness.}
  \label{tab:2photon_GALEX}
\end{deluxetable}

The \GALEX\ FUV band\footnote{The blue edge was chosen to avoid
Ly$\alpha$ from Earth and the Interplanetary Medium as well as the bright
airglow O~I $\lambda1304\,$\AA\ triplet. These lines are bright:
as seen by HST, Ly$\alpha$ is 2\,k$R$ \citep{A12} while the O~I line
13\,$R$, even in Earth's deep shadow.  } is formally
1350--1750\,\AA\ while the NUV band is 1750--2800\,\AA\ (see
Table~\ref{tab:2photon_GALEX}).  At one end of the probability
distribution function (Equation~\ref{eq:Ay}), $y\rightarrow 1/2$,
the two photons have equal energy in which case the NUV band will
register two photons. At the other end, $y\rightarrow 0$, one photon
will get registered in the FUV band and the other will be in the
optical/infra-red (OIR) band.  Integrating over $A(y)$  we find
that each decay results in 0.42 photons in the 1350--1750\,\AA\
band, 0.67 photons in the 1750--2800\,\AA\ band and 0.82 in the OIR
band. [If the division is done by energy then the corresponding
fractions are 33\%, 37\% and 30\%].

\citet{mcb+07} provide a summary of the instrumental
parameters\footnote{The tables for the effective area for FUV and
NUV detectors were obtained from
\url{https://asd.gsfc.nasa.gov/archive/galex/tools/Resolution_Response/index.html}}
as a function of $\lambda$ but provide a single FoV value, $\Omega$,
for each of the two channels.    For each channel, the
band-pass weighted quantities for an input two-photon spectrum, is
computed as follows,
 \begin{equation}
	j_\lambda= \frac{1}{\Omega\mathcal{I}}\int j_\lambda A_{\rm
	eff}(\lambda)d\lambda, \qquad n(2\gamma)=\frac{\Delta\lambda_{\rm
	eq}}{\mathcal{I}}\int \lambda\frac{j_\lambda}{hc} A_{\rm
	eff}(\lambda)d\lambda
  \label{eq:jnu_nnu}
 \end{equation}
where $\mathcal{I}$, the area-bandwidth product and $\Delta\lambda_{\rm
eq}$, the equivalent width, are defined in Table~\ref{tab:2photon_GALEX}.
The unit for $n(2\gamma)$ (listed in Table~\ref{tab:2photon_GALEX})
is count, resulting from our  choice of 1\,sr,  an area of 1\,cm$^2$
and an integration time of 1\,s.  As can be seen from
Table~\ref{tab:2photon_GALEX}, each decay results in 0.27 photon
in the FUV band and 0.44 photon in the NUV band. Consider an
astronomical object emitting only via the two-photon process.
Integrating over the pixels of this object's image, the ratio of
FUV to NUV count rates is expected to be about 0.37.  We end this
section by noting \citet{bba+20} undertook a similar exercise but
with a somewhat different approach.  The two results agree to better
than 10\%.

\section{Atomic Physics}

\subsection{Mixing of 2s-2p states by collisions}
 \label{sec:2s_2p_collisional_mixing}

Recall that an H atom remains, on an average, in the  2s state for
a duration of $A_{21}^{-1}$ or about 0.12\,s. This is sufficiently
long that over this period the atom could get perturbed.
\citet{P52} showed that distant encounters with protons are more
effective than electrons (which had been studied in earlier literature)
in changing the angular momenta of H atoms.  The computed collisional
coefficients are given in Table~\ref{tab:collisions_2photon}.  The
resulting critical density, $\approx 10^4\,{\rm cm^{-3}}$,
significantly reduces the signal of the corresponding
fine structure radio transitions from prime targets such the Sun
or even HII regions \citep{dtm05}.

 \begin{deluxetable}{rrrr}[hbt]
 \tablecaption{Collisional Coefficients for H~I $2s\rightarrow 2p$}
 \tablewidth{0pt}
 \tablehead{
 \colhead{transition} &
 \colhead{collider} &
 \colhead{$q_0\ ({\rm cm^3\,s^{-1}})$} &
 \colhead{$\gamma$}
 }
\startdata
 $2s\ ^{2}S_{\sfrac{1}{2}}\rightarrow 2p\ ^{2}P_{\sfrac{1}{2}}$ & $p^+$ & $2.51\times 10^{-4}$ & $-0.27$\\
 $2s\ ^{2}S_{\sfrac{1}{2}}\rightarrow 2p\ ^{2}P_{\sfrac{3}{2}}$ &     "       & $2.23\times 10^{-4}$ & $-0.03$\\
  $2s\ ^{2}S_{\sfrac{1}{2}}\rightarrow 2p\ ^{2}P_{\sfrac{1}{2}}$ & $e^-$ &$0.22\times 10^{-4}$ & $ -0.37$ \\
  $2s\ ^{2}S_{\sfrac{1}{2}}\rightarrow 2p\ ^{2}P_{\sfrac{3}{2}}$ &     "      &$0.35\times 10^{-4}$ & $ -0.37$\\
 \enddata
\tablecomments{Collisional coefficient at temperature $T=10^4T_4\,$k
 is given by $q=q_0T_4^\gamma$. Condensed from \citet{O74}.}
 \label{tab:collisions_2photon}
\end{deluxetable}

\subsection{Excitation of 2s level by solar photons}
 \label{sec:Excitation_2s_Solar}
 
\citet{bhn+01} briefly mention the excitation of H atoms in
2s state by solar Balmer photons. Here, in addition, we consider bound-free
ionization.

\noindent{\bf Excitation by H$\alpha$ and H$\beta$ (bound-bound).}
An H atom in the ${\rm 2s}\ ^2S_{\sfrac{1}{2}}$ state (level
degeneracy, $g_l=2$) can be excited by bound-bound process, for
instance by absorption of a solar H$\alpha$ photon. The excitation
is then either to ${\rm 3p}\ ^2P_{\sfrac{1}{2}}$  ($g_u=2$) or ${\rm
3p}\ ^2P_{\sfrac{3}{2}}$ level ($g_u=4$). The oscillator strength,
$f_{lu}\propto A_{ul}g_u/g_l$  and  the oscillator strengths,
summed over the two transitions, is $f_{2s\rightarrow 3p}=0.4360$.

\begin{figure}[htbp]    %Solar_Halpha.m
  \centering \includegraphics[width=3.5in]{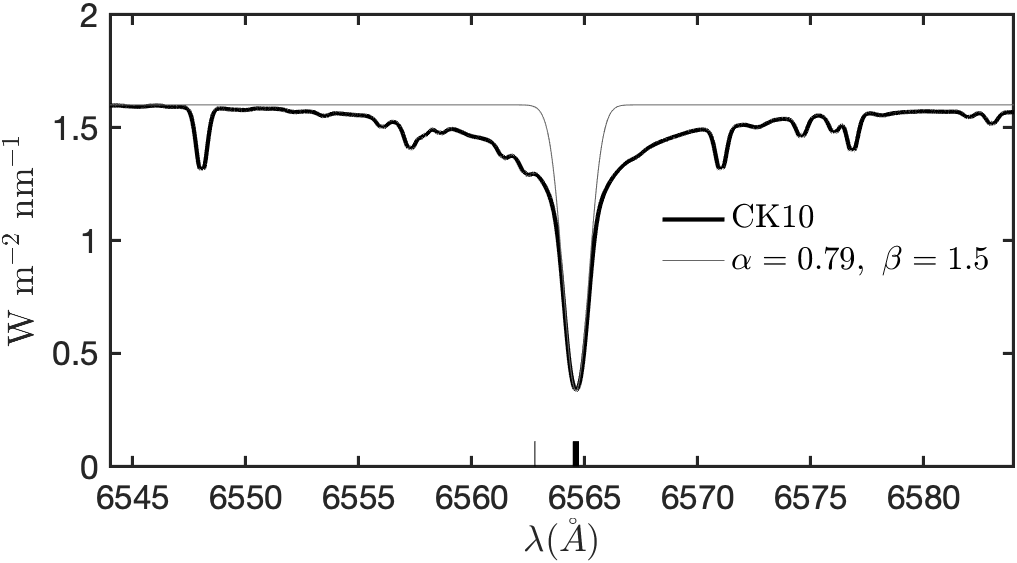}
   \caption{\small High-resolution spectrum  of the Sun in the
   vicinity of H$\alpha$ (data  from \citealt{ck10}).  The absolute
   level at 6500\,\AA\ agrees to within a percent of the spectrum
   from SOLSPEC, aboard the International Space Station \citep{msd+21}.
   The dotted curve is a ``chi-by-eye" fit restricted to the narrow
   H$\alpha$ core.  The model is given by
   $y(\lambda)=1.6\times[1-\alpha\exp(-\beta x^2)]$ where
   $x=\lambda-\lambda(H\alpha)$, $\alpha=0.79$ and
   $\beta=1.5\,\text{\AA}^{-2}$.  The  thick vertical stub is the
   vacuum wavelength, $\lambda(H\alpha)=6564.614$\,\AA.  The thin
   stub is the rest air wavelength of H$\alpha$ (6562.801\,\AA),
   shown merely for reference.}
 \label{fig:SolarHalpha_HighRes}
\end{figure}

The thermal velocity dispersion of H atoms is $\sigma_v=\sqrt{kT/m_H}$
or $2.9\,{\rm km\,s^{-1}}$ for $T=1,000\,$K; here, $m_H$ is the
mass of an H atom. The corresponding FWHM is ${\rm
ln}(256)^{\sfrac{1}{2}}\sigma_v\approx 6.8\,{\rm km\,s^{-1}}$.  We
fit the solar H$\alpha$ absorption feature to a continuum+Gaussian
absorption model (Figure~\ref{fig:SolarHalpha_HighRes}) and derive
an FWHM of $2\sqrt{{\rm ln}(2\alpha)/\beta}=46\,{\rm km\,s^{-1}}$.
Since the geo-coronal H atoms have  little Doppler shift with respect
to the Sun we can assume that the flux of the solar H$\alpha$ line
is constant over the thermal frequency spread of an H atom.  The
H$\alpha$ flux at the bottom of the absorption line is ${\rm
0.34\,W\,m^{-2}\,nm^{-1}}$ (Figure~\ref{fig:SolarHalpha_HighRes}).
This corresponds to $F_{\lambda}(0)=1.1\times 10^{13}\,{\rm
phot\,cm^{-2}\,s^{-1}}\,\text{\AA}^{-1}$ or $F_\nu(0)= 158\,{\rm
phot\,cm^{-2}\,s^{-1}\,{\rm Hz}^{-1}}$.  Then the rate of H$\alpha$
pumping, per atom, is
 \begin{equation}
  \mathcal{R}_{H\alpha}= F_\nu(0)\frac{\pi e^2}{m_e c} f_{2s\rightarrow
  3p}\approx 1.8 d_{\rm AU}^{-2}\ {\rm  atom^{-1}\,s^{-1}}
     \label{eq:RHalpha}
 \end{equation}
where $d$ is the distance in AU.  A similar exercise, carried out
for solar H$\beta$ absorption, yields
 \begin{equation} R_{H\beta} \approx 0.27 d_{\rm AU}^{-2}\ {\rm
 atom^{-1}\,s^{-1}}.
  \label{eq:RHbeta}
 \end{equation}
The reduction is due to  a smaller oscillator strength, $f_{2s\rightarrow
4p}=0.1028$ and a slightly smaller solar spectral flux, $F(0)=104\,{\rm
phot\,cm^{-2}\,s^{-1}\,Hz^{-1}}$. 

\begin{figure}[htbp]	%photoionization_2s.m
 \centering
  \includegraphics[width=3.5in]{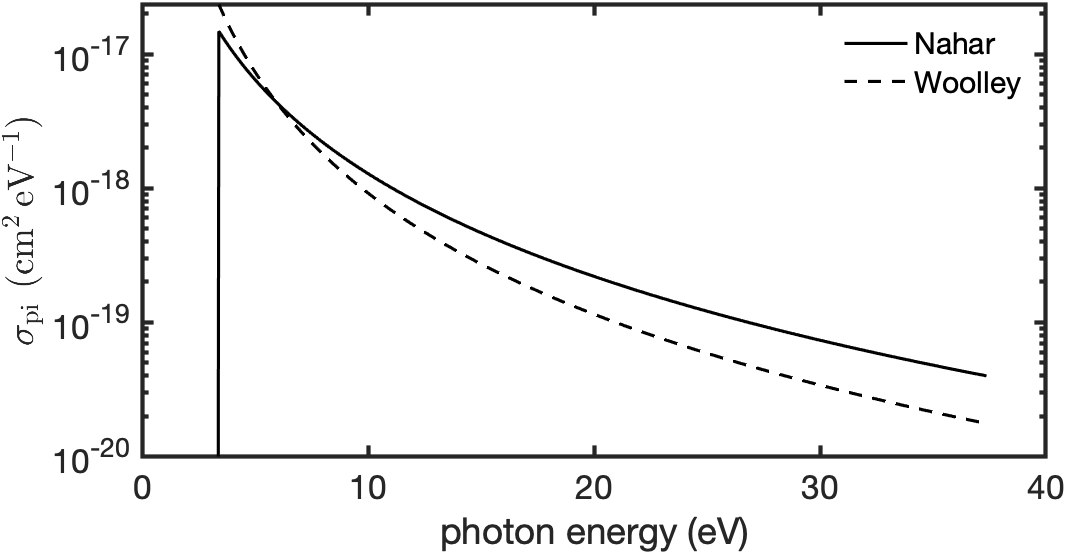}
   \caption{\small 
The run of bound-free cross-section
   as a function of energy of photon (in eV) for the 2s state of
   hydrogen from \citet{W34} and \cite{N21}.  The cross-section
   vanishes for photon energy below the ionization edge, $I_H/4$
   where $I_H$ is the ionization energy of hydrogen. }
 \label{fig:Bound_free_2s}
\end{figure}

\noindent{\bf Bound-free excitation.} 
The excitation
rate from the  2s state\footnote{For 2p levels optical pumping by
an external source cannot compete with spontaneous decay to the
ground state.} to the continuum is given by $ \mathcal{R}_{\rm bf}
= \int_{\nu_\alpha}^\infty\sigma_{\rm bf}({\nu;\rm 2s})F(\nu)/(h\nu)d\nu$
where  $F(\nu)$ is the solar spectral flux and $\sigma_{\rm bf}(\nu;
{\rm 2s})$ is the bound-free cross-section.

\citet{W34} offers a simple method to compute this cross-section
for $\sigma_{\rm bf}(\nu; {\rm 2s})$.  Let $f_c$ be the oscillator
strength for this transition. Then $\sigma_{\rm bf}(\nu; {\rm
2s})=(\pi e^2)/(m_ec)f_c\phi_\nu$ where $\phi_\nu$ is the normalized
frequency distribution. The bound-free cross-section vanishes for
photon energy ($h\nu_0$) below ionization potential from the 2s
state, $I_H/4$ where $I_H$ is the ionization potential of hydrogen.
It is reasonable to assume that the functional form of the bound-free
cross-section is $\propto \nu^{-3}$. The normalization requirement,
$\int_{\nu_0}^\infty\phi_\nu d_\nu=1$, then yields
$\phi_\nu=2\nu_0^{-1}(\nu/\nu_0)^{-3}$.  Next, per the Thomas-Reiche-Kuhn sum
rule, $f_c+\sum_{n=3}^\infty f(2{\rm s}\rightarrow n{\rm p})=1$;
here $f(2{\rm s}\rightarrow n{\rm p})$ is the oscillator strength
for 2s-$n$p transition.  Using the oscillator strengths for the
Balmer series\footnote{At little peril, we ignore  two-photon decay.}
we find $f_c=0.362$. Thus, $\sigma_{\rm bf}(\nu; 2{\rm s})=1.3\times
10^{28}\nu^{-3}\, {\rm cm^2\,Hz^{-1}}$.  Alternatively, we can use
modern calculations for $\sigma_{\rm bf}(\nu)$ \citep{N21}.  The
run of $\sigma_{\rm bf}(\nu)$ with frequency, $\nu$, for the two
approaches is shown in Figure~\ref{fig:Bound_free_2s}. The rate
of photo-ionization by solar photons from the 2s state, using either
formulae, is essentially the same:
 \begin{equation} %supplied by "photoionization_2s.m"
  \mathcal{R}_{\rm bf} = \int_{\nu_\alpha}^\infty\sigma_{\rm bf}(\nu;
  {\rm 2s})\frac{F(\nu)}{h\nu}d\nu
	=0.125 d_{\rm AU}^{-2}\ {\rm atom^{-1}\,s^{-1}}.
  \label{eq:Rbf}
 \end{equation}
Here, we used SOLSPEC data for $F(\nu)$  (see
Figure~\ref{fig:SolarHalpha_HighRes}). Incidentally, the resulting
bound-free ionization rate is a tenth of ionization of H atoms by
the Lyman continuum photons.
The overall rate  of solar photonic excitation of  2s state  is the
sum of the rates given by Equation~\ref{eq:RHalpha}, \ref{eq:RHbeta}
and and \ref{eq:Rbf}, $A_p\approx 2.2d_{\rm AU}^{-2}\,{\rm
atom\,s^{-1}}$.  Solar optical pumping reduces the production of
two-photon continuum by $A_{2\gamma}/ (A_{2\gamma}+A_p)$.

\begin{figure}[htbp]	%OxygenGrotrian.m, Oxygen_Lymanbeta.m 
 \centering
  \includegraphics[height=2in]{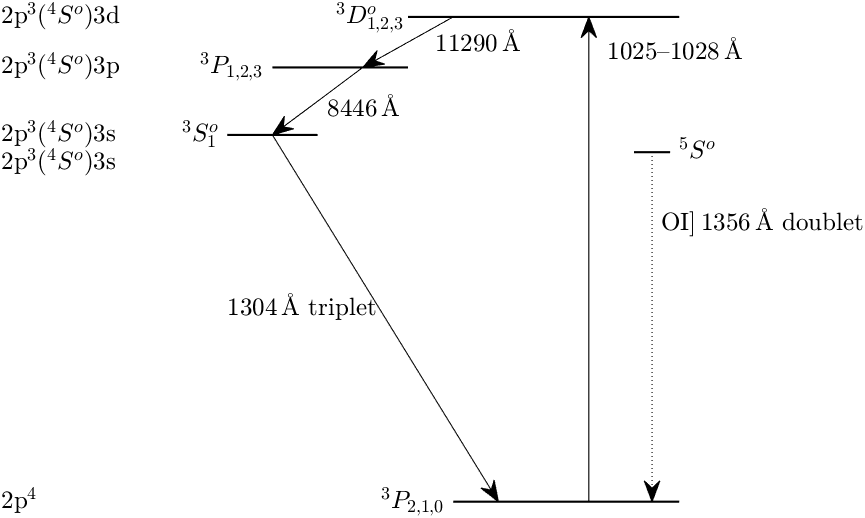}\qquad\qquad
   \includegraphics[height=2in]{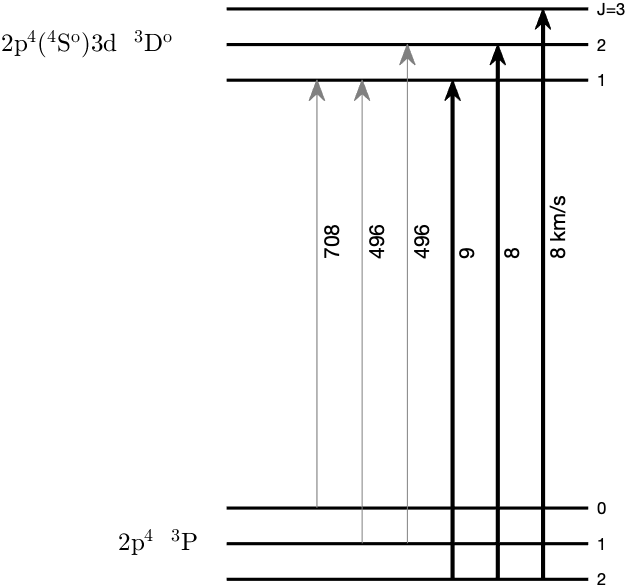}
    \caption{\small ({\it Left}). Partial Grotrian diagram for O~I
    (to scale) displaying key allowed transitions.  Solar Ly$\beta$
    photons excite O~I atoms from the ground state, ${\rm 1s^22s^22p^4}\
    ^3P$, to the ${\rm 1s^22s^22p^3}(^4S^0){\rm 3d}\ ^3D^o$ excited
    state. The O~I atom can decay back to the ground state or decay
    to  the ${\rm 1s^2s2^22p^3}(^4S^o){\rm 3p}\  ^3P$ state which
    then decays to ground state emitting along the way
    the famous O~I\,1304\,\AA\ triplet (composed of
    $\lambda\,1302.17\,$\AA, $\lambda\,1304.86\,$\AA,
    $\lambda\,1306.03\,$\AA\ lines).   The airglow O~I]\ $\lambda\lambda
    1355.56$\AA, 1358.51\,\AA\ is a spin-forbidden ($\Delta S\neq
    0$; ``inter-combination", ``semi-forbidden") transition and
    results from electron excitation of O~I from the ground state
    to the $2{\rm p}^3(^4S^o)3{\rm s}\,^5S^o$ level 
    (transition shown by dotted line). ({\it Right}): Grotrian
    diagram of O~I (not to scale) restricted to the six allowed
    transitions between the ground state and ${\rm 1s^22s^22p^3}(^4S^0){\rm
    3d}\ ^3D^o$.  The wavelength for each transition is converted
    to a velocity w.r.t. the rest wavelength of Ly$\beta$. The
    rightmost three lines (thick black color) have sufficiently
    small velocity shifts, $\approx 9\,{\rm km\,s^{-1}}$, that they
    can be  excited by solar Ly$\beta$ photons.  These three lines
    are referred to as the ``trio" in the main text.  The remaining
    three (gray) lines have large velocity shifts, 500--700\,km\,s$^{-1}$,
    and so cannot be excited by solar Ly$\beta$.}
 \label{fig:Oxygen_Bowen}
\end{figure}

\subsection{Bowen Fluorescence of O~I by Ly$\beta$}
 \label{sec:OI_BowenFluorescence}
 
The reason that the measurements in Table~\ref{tab:STP_78} are
listed as Ly$\beta$+OI is because there happens to be a near
coincidence between Ly$\beta$ and an excited state of O~I, ${\rm
1s^22s^22p^3}(^4S^o){\rm3d}\ ^3D^o$ (\citealt{map+87}; see
Figure~\ref{fig:Oxygen_Bowen} for a partial Grotrian diagram of
O~I).  The oscillator strengths  for all the transitions can be found
from the NIST database.  The excited OI atom has a probability of
71\% to return to the ground state and 29\%  probability of decaying
to the ${\rm 1s^22s^22p^3}(^4S^o){\rm 3p}\ ^3P$ state with  subsequent
cascade to ground state with the last lap involving the famous O~I
$\lambda 1304\,$\AA\ triplet.  In addition to this channel, oxygen
atoms in the thermosphere are directly excited by the solar
chromospheric O~I $\lambda 1304\,$\AA\ triplet (in emission).
Together, these two processes account for the brilliance of the
airglow O~I triplet.  In contrast, the other FUV line of oxygen,
O~I] $\lambda\lambda 1356, 1358\,$\AA, is weaker because it is
spin-forbidden and is primarily excited by collisions with electrons.
For satellites in LEO, such as HST,  even in Earth's
shadow, the airglow is dominated by Ly$\alpha$ (2\,kilo-Rayleigh
or $2\,{\rm k}R$), the O~I $\lambda\lambda$1302\,\AA\ triplet
(1\,k$R$) and the semi-forbidden O~I] $\lambda\lambda$ 1356\,\AA\
doublet ($0.1\,{\rm k}R$.  \GALEX, like HST, was in LEO.  The FUV
band of \GALEX\ was carefully chosen to exclude Ly$\alpha$ and the
strong O~I triplet and attenuate the semi-forbidden O~I] line
\citep{mfs+05}.

\section{The Thermosphere}
 \label{sec:Thermosphere_Appendix}
 
The troposphere (0--12\,km) contains most of the atmosphere.  The
temperature within the stratosphere (12--50\,km) increases with
height owing to absorption of solar UV by ozone.  Once the ozone
is dissociated temperature starts to decrease again with height
until dissociation of oxygen and other molecules begins in the
thermosphere.   The mesosphere (50--85\,km) is the layer between
the stratosphere and the thermosphere.  It is in this layer in which
meteors burn, providing a convenient layer of Na~I for laser
guide-star adaptive astronomy.

\begin{figure}[htbp]    %ThermosphereProfile.m
 \centering
  \includegraphics[height=2.5in]{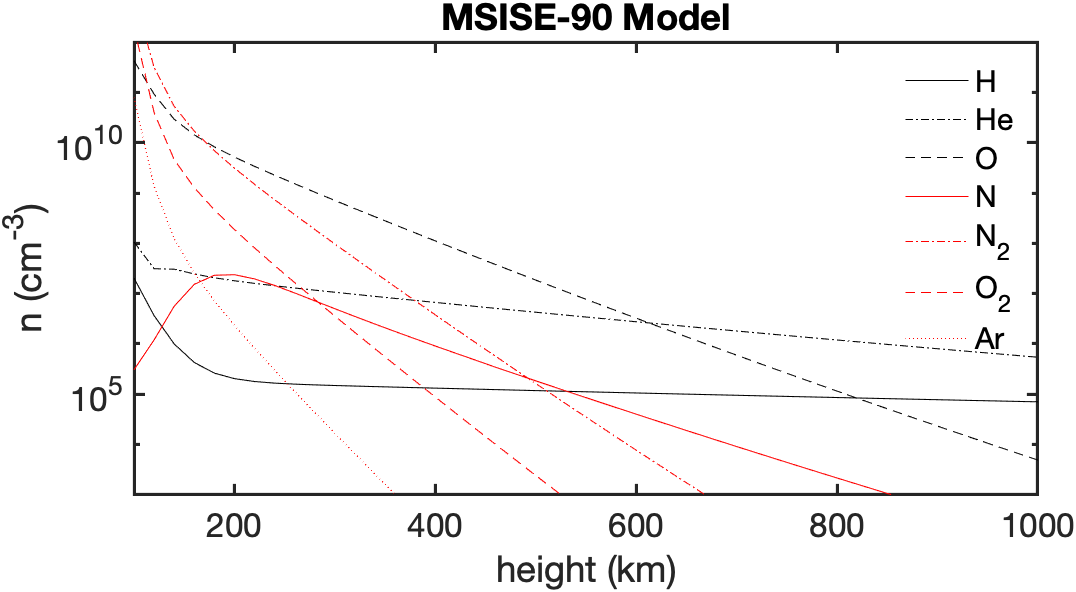} \ \
   \includegraphics[height=2.35in]{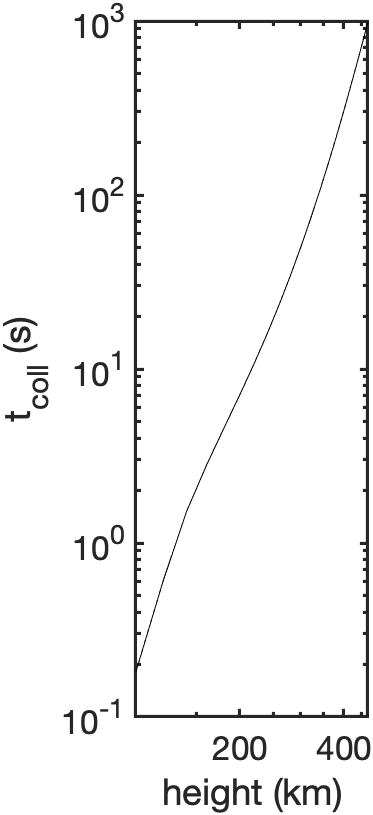}
    \caption{\small (Left) MSIS-E-90 atmosphere model for latitude
    of 30$^\circ$ and longitude of 0$^\circ$ on 01-January-2000 at
    a local time of 0100.  From
    \url{https://ccmc.gsfc.nasa.gov/modelweb/models/msis_vitmo.php}
    (Right). The run, with height, of the mean time between collisions
    between an H atom and a neutral atom, $\tau=(\sigma_0\sum_i
    n_iv_i)^{-1}$ where $\sigma_0=10^{-16}\,{\rm cm^{2}}$ is the
    characteristic ``hard sphere" approximation for the cross section
    and $n_i$ is the number density of species $i$ and $v_i=\sqrt{3kT/m_H}$
    is the relative velocity between H atom and species $i$. }
 \label{fig:Thermosphere_Profile}
\end{figure}

The thermosphere (80 to about 700\,km) is the region in which low-earth
satellites are located (e.g., International Space Station: 420\,km;
HST 540\,km; Swift Observatory: 550\,km; {\it GALEX}: 685\,km; FUSE,
750\,km).  The temperature in the thermosphere increases with height,
owing to the absorption of solar EUV radiation by majority molecular
species.  Thus atoms  are the primary constituents of the thermosphere.
The temperature  is strongly dependent on the solar EUV irradiance,
ranging from 800\,K to 2000\,K (with strong day/night dependence).
The radial extent of the thermosphere is quite sensitive to solar
EUV, puffing up to 1000\,km during solar maximum and receding to
500\,km during solar minimum.

\begin{figure}[htbp]
 \centering
  \includegraphics[width=3in]{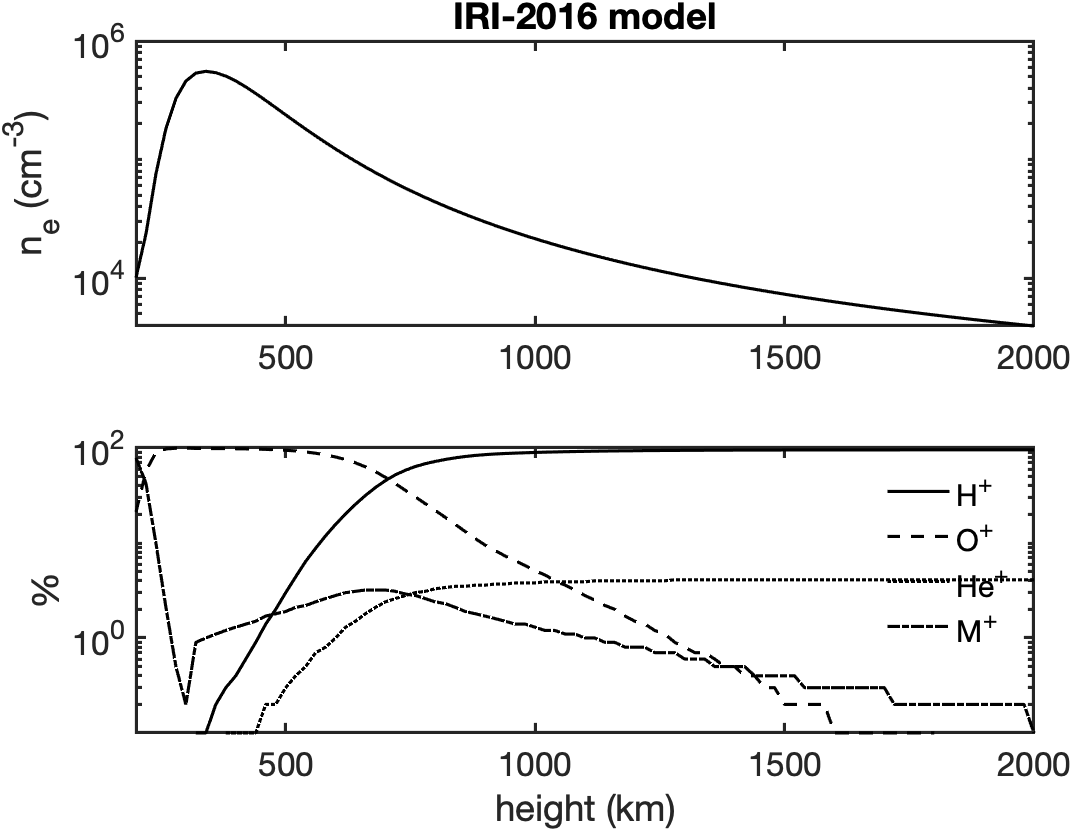}
   \caption{\small (Top) The run of electrons with altitude. (Bottom)
   The  run of H$^+$, He$^+$, O$^+$ and other ions (``M$^+$":
   O$_2^+$, N$^+$ and NO$^+$), but expressed as a percentage of the
   electron density.  }
 \label{fig:Electron_Profile}
\end{figure}

The density, ionization fraction and temperature of the thermosphere
were obtained from the Community Coordinating Modeling Center (CCMC)
portal\footnote{ \url{https://ccmc.gsfc.nasa.gov/about.php}}:
MSIS-E-90 (``Mass Spectrometer and Incoherent Scatter radar -
Exosphere-[19]90") for the run of neutral particles
(Figure~\ref{fig:Thermosphere_Profile}) and IRI-2016 (International
Reference Ionosphere -- 2016). Hydrogen is a minority species in
the thermosphere and, furthermore, suffers from outflow. As such,
\citet{bhn+01} argue that the hydrogen density profile in MSIS
is not reliable.

\begin{figure}[htbp] 
 \centering
  \includegraphics[width=2.3in]{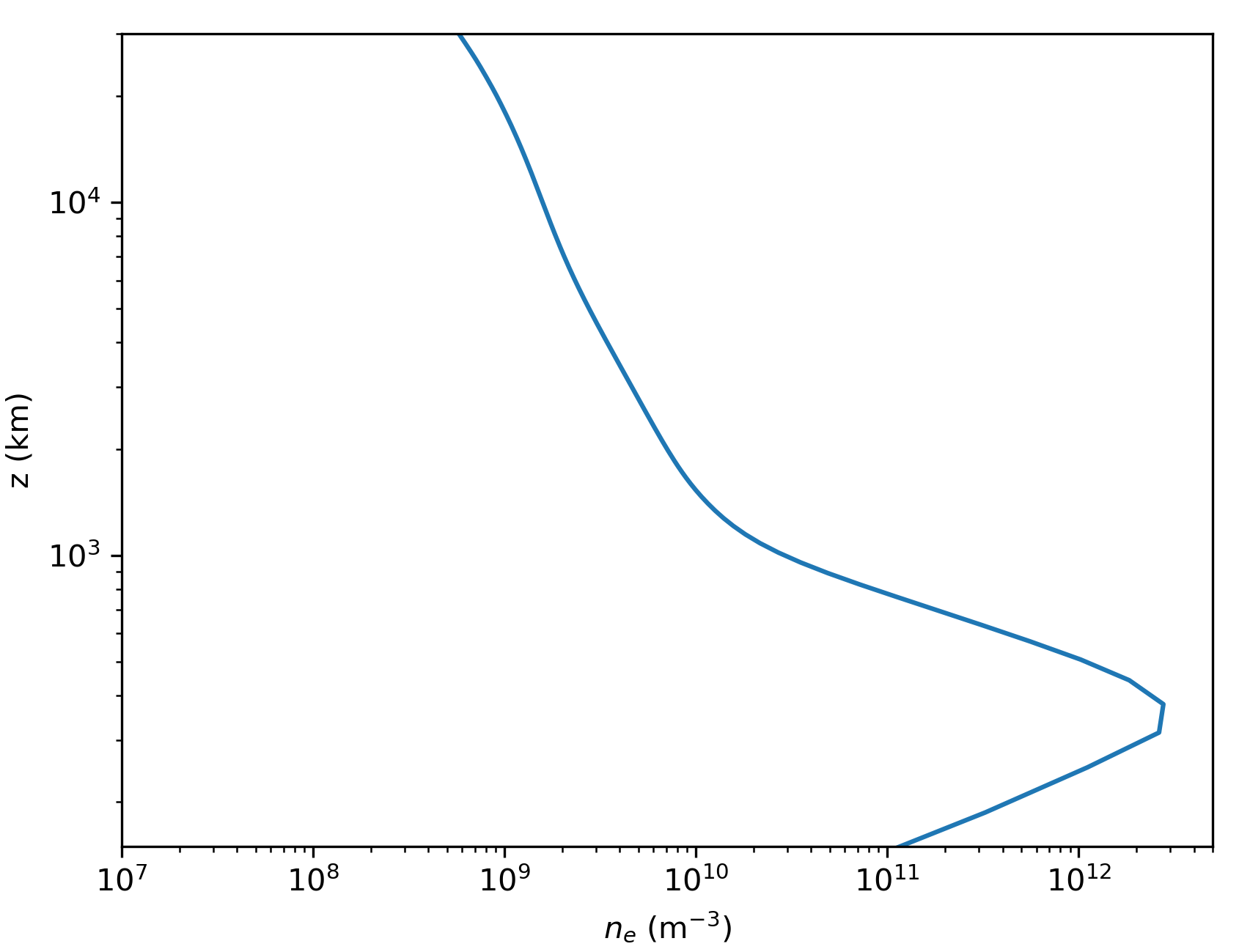}
   \caption{\small Radial profile of electrons (equatorial plane).
   Note the unit for electron density is m$^{-3}$.  Figure supplied
   by Matthew D. Zettergren.  }
 \label{fig:Extended_Ionosphere}
\end{figure}

\section{The Ionosphere}
 \label{sec:TheIonosphere}
 
As can be seen from Figure~\ref{fig:Electron_Profile} the IRI-2016
model for electrons stops at 2,000\,km.  The higher altitude profile
was generated, at my request, by Matthew Zettergren, Embry-Riddle
Aeronautical University. The GEMINI open-source ionospheric
model\footnote{\url{https://github.com/gemini3d}} was used in a 2-D
meridional, dipole configuration to simulate plasma density evolution
over several days (e.g., \citealt{zs15}).  The grid used covers
$\pm 60^\circ$ in latitude, corresponding to apex altitudes of about
32,000\,km (altitude of the model grid above magnetic equator).
GEMINI solves conservation of mass, momentum, and energy equations
for the ionospheric plasma for 6 ion species relevant to the
terrestrial ionosphere, including protons.  The model was run
moderate for high solar and geomagnetic activity levels of F10.7=129.5,
F10.7a=104.7, and solar index Ap=37.  The date of the simulation
is 10/6/2011 (near equinox) and the UT is about 5:45 (corresponding
roughly to noon local time), representing a typical daytime plasma
density state.  Figure~\ref{fig:Extended_Ionosphere} shows a profile
extracted from the geomagnetic equator.  The results are meant to
be illustrative of plasmasphere conditions during geomagnetically
quiet times.

\section{The Exosphere}
 \label{sec:Exosphere_Appendix}
 
The exosphere is defined as the region in which the collisions of
neutral particles with each other ceases to be important. The base
of the exosphere (``exobase") depends on solar activity but a typical
value is 500\,km. Three families of particles are defined as follows:
``ballistic" -- particles that lack sufficient speed and so fall
back; ``escapers" -- particles which have sufficient speed to escape;
and ``satellite" --- particles which undergo a rare collision
($R\lesssim 2.5\,R_E$) which sends them back down.  Thus the particle
density of the exosphere is not a simple power law.  The temperature
in the exosphere decreases to one third of the base value at $4\,R_E$
and two-fifth at $10\,R_E$; here, $R_E$ is the radius of Earth
($\approx 6,400\,$km). We adopt the ``standard'' temperature  of
1025\,K (cf.\ \citealt{omf+03}). The corresponding thermal rms
velocity is 2.9\,km\,s$^{-1}$.

\section{STP\,78-1 \&\ Minisat-01}
 \label{sec:LEO-satellites}
 
The Space Test Program 78-1 (STP\,78-1; aka
``Solwind")\footnote{Unfortunately, towards the end of the mission,
Solwind was assigned as target for a pilot demonstration of
anti-satellite technology (ASAT).  On September 13, 1985 the satellite
was destroyed by an ASM-135 missile mounted on an F-15 fighter
airplane.} was launched in 1979 into a sun-synchronous orbit (600\,km
height). The satellite spin-orbital axis was perpendicular to the
Earth-Sun axis (i.e., a ``noon-midnight" orbit).  It carried, amongst
other instruments, an EUV/FUV spectrometer for studies of airglow
\citep{ckb84}. The spectrometer had an entrance window of $18^\circ\times
9^\circ$ and operated over 350--1400\,\AA\ band.  \citet{ckb84}
reported  satellite night-time EUV spectrum of airglow, both looking
``down" (zenith angle, $z$, between 120$^\circ$ and 150$^\circ$)
and looking ``up'' (zenith angle between $30^\circ<z<80^\circ$).\footnote{
We interpret the geometry as follows: the line starting at the Sun,
going through Earth and then the satellite defines the axis with
which $z$ is measured; thus, $z=0$ is when the satellite is in the
night sky and the Earth is below the satellite.} The relevant
measurements are summarized in Table~\ref{tab:STP_78}.

The Spanish Minisat-01 spacecraft \citep{mgt+98} carried a high
spectral resolution EUV spectrometer, EURD (``Espectr\'ografo
Ultravioleta extremo para la Radiaci\'on Difusa"; \citealt{eka+06}).
The orbit was a circle with height of 580\,km and inclined 151$^\circ$
with respect to the equator.  Observations were obtained only during
satellite midnight, specifically restricted to zenith angle of
$-85^\circ$ (just before ground dawn)\footnote{In contrast to the
STP-78 convention the authors appear
to have assigned a sign to $z$ depending on whether line-of-sight
is towards East or West} and $+80^\circ$ (just after dusk).
Restricting to absolute zenith angle of $<70^\circ$, Ly$\beta$ was
detected at a level of $6.4\,R$ \citep{lmg+01}.

\section{SWAN \&\ IMAGE}
 \label{sec:SWAN-IMAGE}

The Solar \&\ Heliospheric Observatory (SOHO) is a ESA-NASA mission
that is located in the vicinity of the Earth-Sun L1 region and
focused on the studies of the atmosphere of the Sun, the solar wind
and helio-seismology. It carries Solar Wind ANistoropies (SWAN)
instrument whose primary goal is to  study the structure of the
solar wind through its interaction with the IPM.  A hydrogen cell
acts by absorbing the incident solar light at the rest wavelength
of Ly$\alpha$.  In effect, the cell provides a spectral resolution
of $10^5$ \citep{bkq+95}. SWAN data has also been used to study the
distribution of geo-coronal H atoms \citep{bbq+19}.

IMAGE (Imager for Magnetopause-to-Aurora Global Exploration) was a
NASA Medium Explorer class mission that was designed to study the
response of Earth's magnetosphere to changes in the solar wind. It
was launched into a highly elliptical orbit (1,000\,km$\times$46,000\,km)
with an inclination of 90.01$^\circ$ and an orbital period of about
14\,hours. Its payload included an FUV imaging system which included
the ``GEO" photometer \citep{mhf+00}.  The three photometers,
oriented differently, respond to radiation coming from within their
1-degree FoV in the wavelength range 1150--1500\,\AA.  This instrument
was designed to measure the brightness of the geo-coronal Ly$\alpha$.
\citet{omf+03} provide a two-exponential model fit to the observed
brightness. An example fit is
 \begin{equation}
  I(r) = 16.9\exp(-r/1.03) + 1.06\exp(-r/8.25)\,{\rm k}R
 \end{equation}
where $I(r)$ is the Ly$\alpha$ intensity  and the radius $r$ is in
units of $R_E$. Under the (admittedly simplistic) assumption of the
medium being optically thin the authors invert the observations and
provide density profile for H atoms.

\end{document}